\documentclass[journal]{IEEEtran}
\usepackage{latexsym}
\usepackage{lettrine}
\usepackage{graphicx}
\usepackage{amsfonts,amssymb,amsmath}
\usepackage{hyperref}
\usepackage[utf8]{inputenc}
\usepackage{cite}
\usepackage{floatrow}
\usepackage{amsmath,amssymb,amsfonts,stfloats}
\usepackage{tabularx}
\usepackage{graphicx}
\usepackage{textcomp}
\usepackage{floatrow}
\usepackage{xcolor}
\usepackage{booktabs}

\usepackage{multicol}
\usepackage{multirow}
\usepackage{acronym}
\usepackage{url}
\usepackage{mathtools}
\usepackage{algorithm}
\usepackage{algpseudocode} 
\usepackage{enumitem}
\usepackage[caption=false,font=footnotesize]{subfig}

\def\BibTeX{{\rm B\kern-.05em{\sc i\kern-.025em b}\kern-.08em
    T\kern-.1667em\lower.7ex\hbox{E}\kern-.125emX}}

\def\BibTeX{{\rm B\kern-.05em{\sc i\kern-.025em b}\kern-.08em T\kern-.1667em\lower.7ex\hbox{E}\kern-.125emX}}

\begin{document}

\title{Localizing the Vehicle's Antenna: an Open \\ Problem in 6G Network Sensing}

\author{Francesco Linsalata, Dario Tagliaferri, Luca Rinaldi,  Lorenzo Bezzetto,  Marouan Mizmizi, Davide Scazzoli, Damiano Badini, Christian Mazzucco, Maurizio Magarini, and Umberto Spagnolini
\thanks{F. Linsalata, Dario Tagliaferri, L. Rinaldi, L. Bezzetto,  M. Mizmizi , D. Scazzoli, and M. Magarini are with Dipartimento di Elettronica, Informazione e Bioingegneria, Politecnico di Milano, Via Ponzio 34/5, 20133, Milano
Italy.}
\thanks{D. Badini and C. Mazzucco are with Huawei Technologies, Centro Direzionale Milano 2, Palazzo Verrocchio, 20090, Segrate, Italy}
\thanks{U. Spagnolini is with Dipartimento di Elettronica, Informazione e Bioingegneria, Politecnico di Milano, Via Ponzio 34/5, 20133, Milano Italy, Huawei Technologies as Huawei Industry Chair and Consorzio Nazionale Interuniversitario per le Telecomunicazioni (CNIT), Italy.}}

\maketitle

\begin{abstract} 
Millimeter Waves (mmW) and sub-THz frequencies are the candidate bands for the upcoming Sixth Generation (6G) of communication systems. The use of collimated beams at mmW/sub-THz to compensate for the increased path and penetration loss arises the need for a seamless Beam Management (BM), especially for high mobility scenarios such as the Vehicle-to-Infrastructure (V2I) one. Recent research advances in Integrated Sensing and Communication (ISAC) indicate that equipping the network infrastructure, e.g., the Base Station (BS), with either a stand-alone radar or sensing capabilities using optimized waveforms, represents the killer technology to facilitate the BM. However, radio sensing should accurately localize the Vehicular Equipment (VE)'s antenna, which is not guaranteed in general. Differently, employing side information from VE's onboard positioning sensors might overcome this limitation at the price of an increased control signaling between VE and BS. This paper provides a pragmatic comparison between radar-assisted and position-assisted BM for mmW V2I systems in a typical urban scenario in terms of BM training time and beamforming gain loss due to a wrong BM decision. Simulation results, supported by experimental evidence, show that the point target approximation of a traveling VE does not hold in practical V2I scenarios with radar-equipped BS. Therefore, the true antenna position has a residual uncertainty that is independent of radar's resolution and implies 50\,\% more BM training time on average. Moreover, there is not a winning technology for BM between BS-mounted radar and VE's onboard positioning systems. They provide complementary performance, depending on position, although outperforming blind BM techniques compared to conventional blind methods. Thus, we propose to optimally combine radar and positioning information in a multi-technology integrated BM solution.  
\end{abstract}


\section{Introduction}\label{sec:intro}

The Sixth Generation (6G) of cellular communications is expected to support advanced mobile and vehicular services achieving unprecedented performance~\cite{SaaBenChe:J20}. To guarantee these requirements, the 3rd Generation Partnership Project (3GPP) has recently established the use of Millimeter Waves (mmW) and sub-THz bands~\cite{TS_38213,ParBlaBlaDahFodGraStaSta:J20}.
Propagation at these frequencies is subject to higher path and penetration losses compared to currently deployed communication systems, calling for the usage of highly directive beams in Multi-Input-Multi-Output (MIMO) systems. The dynamic Beam Management (BM) in high-mobility communications, such as Vehicle-to-Everything (V2X), is however a well known issues at mmW/sub-THz~\cite{CaireBA, surveyBA}. The BM goal is to find the optimal pointing direction for transmitter and receiver maximizing the received power. Conventional BM approaches mostly rely on exhaustive beam search over medium-to-large codebooks~\cite{StoDua:J20,SRIVASTAVA:evc20,SinNanNan:EVC19}, resulting in significant training times and reduced spectral efficiency. Therefore, various solutions aimed at speeding up the BM procedure arose in the literature~\cite{giordani2018tutorial, surveyBA}. Several works leverage the communication interface only, while the most recent and advanced ones exploit side information from heterogeneous sensors, e.g. Global Navigation Satellite System (GNSS) and radar, to seamlessly localize the transmitter and receiver antennas and ease the BM procedure. 

\subsection{BM Literature Review}

The BM approaches can be categorized into two distinct areas: blind methods, which do not exploit the side information, and sensor-assisted methods, which take as input either direct position-related measurements (e.g., from GNSS signalling) or indirect ones, (e.g., by processing radar frames).

\subsection*{Blind methods} 

In 3GPP standards, BM approaches at network infrastructure are supposed to estimate the optimal beam by leveraging only on the communication interface/protocols~\cite{TS_38213}.
The 5G NR standard Initial Access (IA), or BM procedure, is implemented by periodic transmission of Synchronization Signals Block (SSB), which are selectively transmitted by the Base Station (BS) to some specific spatial directions.
The receiving Vehicle Equipmente (VE) searches for the SSBs and by decoding them it is able to synchronize and identify the optimal beam for communicating with the BS. 
The current standard procedure is defined as \textit{blind} method, since it has the advantage of not requiring assistance/side information, such as the position information of VE. Regular BM operations are based on control messages that are periodically exchanged between BS and VEs~\cite{TS_38213}.
However, this procedure can introduce undesired delay as soon as the codebook size increases, thus limiting most of the future 6G requirements in terms of low latency and safe critical applications~\cite{giordani2018tutorial}.
Several works aim at improving the current blind standard procedure without introducing any assistance information. In~\cite{Heath2014Hier_codebook}, the authors developed an adaptive algorithm
to estimate the mmW sparse channel, exploiting/proposing an hierarchical multi-resolution codebook, designed to construct the training beamforming vectors with different beamwidths. 
Variable length codebook detection via the sequential competition and elimination test is proposed in \cite{FetBA} to provide adaptivity to channel conditions, resulting in a rate gain and/or a delay reduction.
To compensate for pointing direction uncertainty, in~\cite{OptimalBeamSweeping}, first, the BS widens the transmission beam until a critical beamwidth, based on a analytical threshold, is achieved. Then, the BS performs a bisection algorithm to refine and tighten the beam. The authors of~\cite{CS_BeamSweeping} formulate the BM process as a sparse problem and use compressive sensing to determine the required number of probed directions. However, all these approaches introduce a significant training time for the channel estimation and/or and algorithm complexity and might not represent a systematic solution to improve the BM. 

\subsection*{Sensor-assisted methods}

\textbf{Sensors at the VE side}: BM approaches leveraging on side information from VE's on board sensors have been recently investigated to improve the perfomances of blind methods~\cite{ba5,ba6,ba7}. In~\cite{ba2}, the BM relies on a fingerprint database containing prior information of favorite candidate pointing directions. The database is queried by means of the knowledge of VE's position within the cell, which is continuously updated. A similar approach is also proposed in~\cite{Ba3} and~\cite{channelMiz}, where the beam selection relies on modal analysis based on the spatial invariance of spatio-temporal Vehicle-to-Infrastructure (V2I) channel modes, computed by exploiting recurrent vehicle passages over the same location in space. The list of these modes is associated with an optimal BS-VE beam pair, to be again queried according to VE's position. For all position-aided BM solutions, GNSS is the most used sensor to provide seamless and updated VE position information, thanks to GNSS penetration in the vehicular market. However, GNSS-only BM solutions suffer from outages and urban canyoning effects, that severely degrade the position accuracy down from few to tens of meters~\cite{Kbayer2018_GPS}. To overcome the latter limitations, recent research trends have focused on either fuse data from multiple sensors on-board the vehicular VE (inertial sensors, wheel odometry, lidars, radars, etc.) or let the vehicles cooperate to improve positioning performance. For instance, the works in~\cite{Brambilla2020,sensorDario} show the benefits of a sensor-assisted BM for V2X links, leveraging on the mutual exchange of transmitter and receiver position and the associated uncertainties, estimated by fusing GNSS and inertial sensors' data. The authors of~\cite{lidarBA}, instead, propose to exploit a lidar, while~\cite{Brambilla2020positioning,mashhadi2021federated} explore the cooperation among vehicles to improve positioning accuracy. Of course, the main drawback of sensor-assisted BM approaches is that VE's position shall be continuously signaled to the BS by means of a proper control channel. Although this is currently considered for vehicular networks in the context of autonomous driving~\cite{Brambilla2020positioning}, the real impact of control overhead is yet to be determined for very dense vehicular networks.

\begin{figure}[!t] 
\centering
\includegraphics[width=0.7\columnwidth]{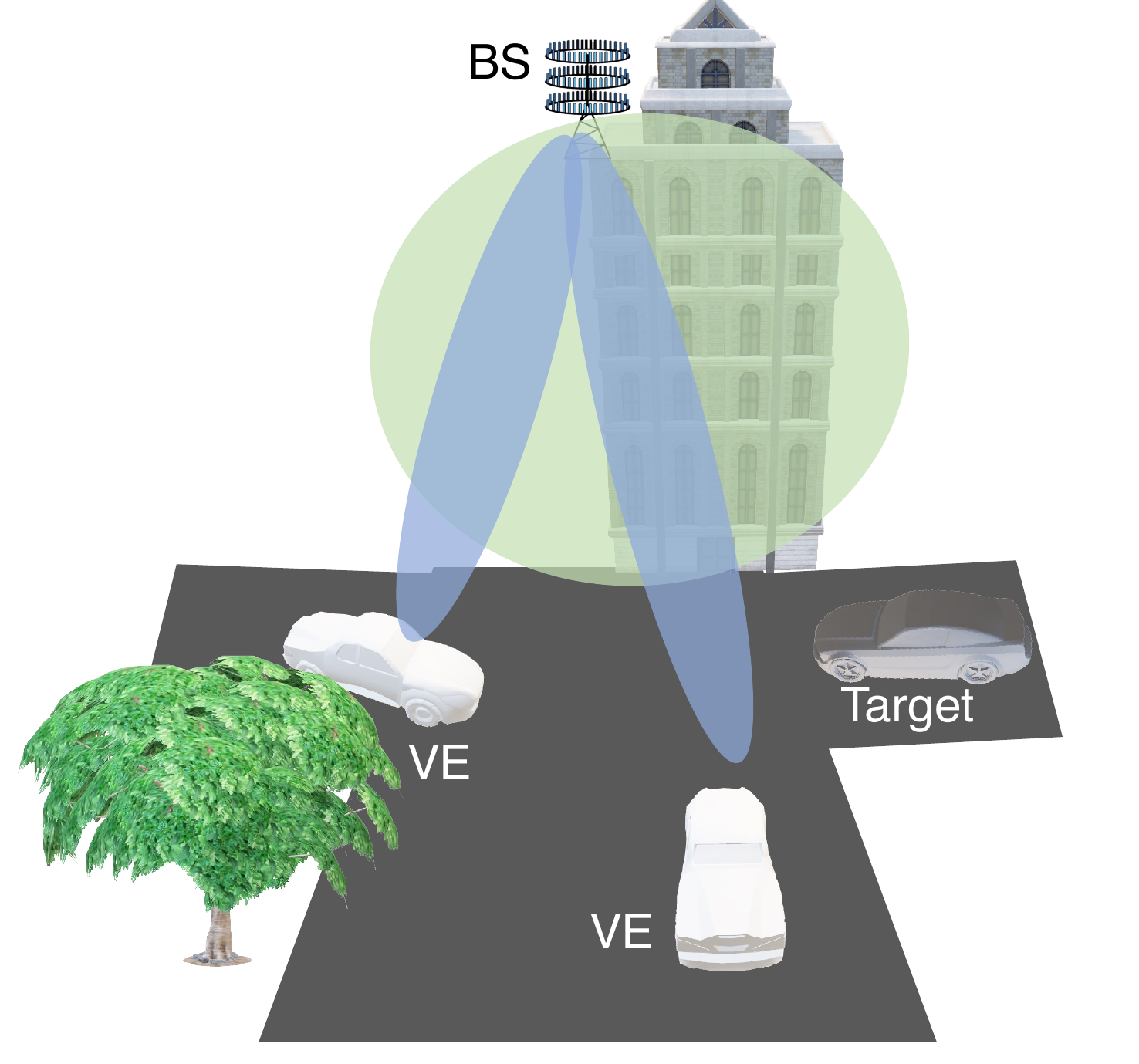}
\caption{Radar-equipped BS, serving VEs (blue beams) while sensing the environment (green beam). }
\label{fig:scenario}
\end{figure}

\textbf{Sensors at the BS side}: In addition to sensors on board the VE, equipping the network infrastructure (e.g. BS) with the capability of \textit{sensing} the environment is the basis for the Perceptive Mobile Network (PMN) concept proposed in~\cite{Heath2020PMN}, motivated by a strong reduction in beam training time~\cite{Heath2020leveragingsensing}. Most of the researches are focused on Integrated Sensing and Communication (ISAC) systems, where both communication and sensing functionalities are merged into a single optimized waveform~\cite{Heath2021overviewJCS}. However, the usage of a radar at the BS, working on a separate frequency band with respect to the communication system, can be considered as an intermediate step toward PMN and represents an upper bound for ISAC in terms of both communication and sensing performance. Radar data can be directly used for BM and blockage prediction.
Radar-aided mmW BM has been investigated for the first time in~\cite{ba8}, where the authors showed the benefits of using a radar for improving the spectral efficiency of mmW V2I links by estimating the V2I channel covariance. The same authors proposed a preliminary comparison of GNSS- and radar-aided BM, recommending the full adoption of the latter approach~\cite{ba11}. However, authors assume a GNSS error radius of $10$ m, which is largely pessimistic even in a dense urban scenario, especially when considering advanced positioning techniques. Moreover, in~\cite{ba11}, the authors assume the perfect knowledge of the beam coherence time~\cite{va2016impact}, thus of VE's velocity, providing an optimistic lower bound on beam training time. Passive radar at the BS is instead investigated in~\cite{ba12,prel2022deep}, exploiting the signals from automotive radars. However, uplink sensing requires the clock synchronization between BS and UE, whose practical implementation is challenging and is not discussed by the authors.
A relevant contribution is given in~\cite{demirhan2021radar}, where the authors experimentally demonstrated the potential of radar-aided beam selection in a typical V2I scenario, where a moving VE is illuminated by a MIMO radar while being served by a fixed mmW unit. The beam selection is performed offline using a deep learning approach directly from range-angle and/or range-angle-velocity radar maps. Results highlight the effectiveness of the proposed approach, even if performance of the radar-aided approach is not compared with other methods. Moreover, all of the aforementioned works assumed the radar capable of detecting the VE position of the antenna, reaching the utter sensing performance for BM. However, the VE behaves like an extended target~\cite{berthold2017scatter}, adding a residual uncertainty on its mmW/sub-THz antenna position that can degrade the BM performance.

 \begin{figure*}[!t] 
\centering
    \includegraphics[width=\columnwidth]{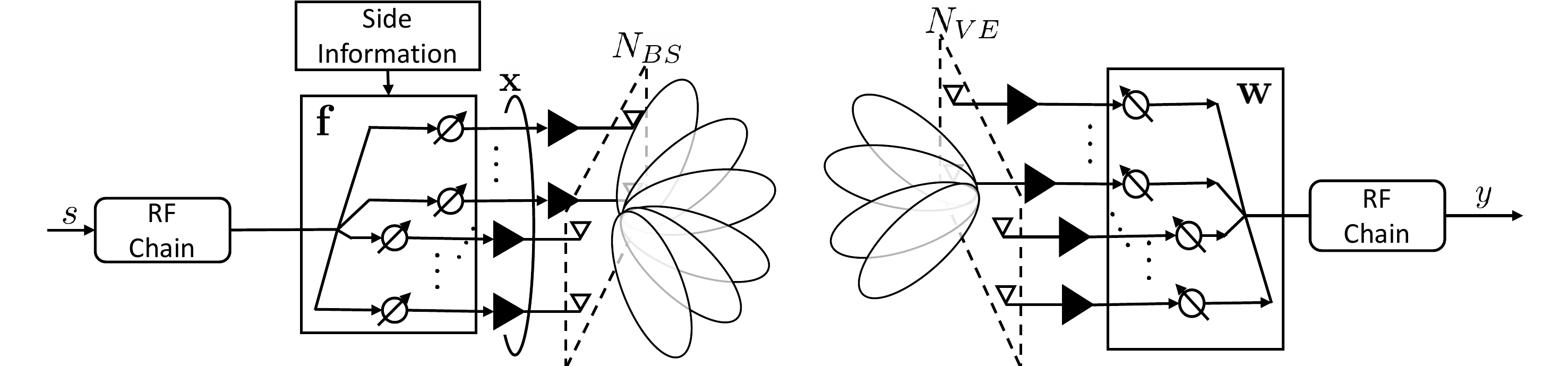}
    \caption{Block diagram of the MIMO communication system between one panel of the BS and the VE. Side information is used for BM.}
    \label{fig:system_model}
\end{figure*}

\subsection{Contributions}
This paper addresses the BM issue in a mmW/sub-THz V2I communication scenario, depicted in Fig. \ref{fig:scenario}, comparing different BM methods.
We advance the current state of the art by proposing the following contributions: 

\begin{itemize} 
\item We compare different BM solutions, both blind and sensor-assisted, in terms of training time and beamforming gain losses. Multiple configurations of each technology are assessed in exhaustive simulations.
Numerical results in different communication regions within the cell determine whether the usage of a BS-mounted radar is enough to provide satisfactory BM performance or not, and in what circumstances the radar is advantageous compared to a positioning system and vice-versa. Overall, sensor-assisted BM solutions are shown to outperform blind BM ones, especially in case of non-continuous communications. Radar at the BS is effective in short-range and boresight regions, with an average beam training time at the BS lower than $1$ ms for each testing. Complementarily, positioning sensors provide better training time performance in medium-to-long-range communications.
Overall, an high-resolution radar (128$\times$4, $64\times 8$ or $32\times 16$ antennas) is comparable with an position systems at the VE characterized by an accuracy lower than $1$ m.
\item According to the evidences from a purposely designed experimental campaign, we show that the VE point target assumption, used in \textit{all} the aforementioned radar works, does not hold for vehicles of medium-to-large sizes (car, trucks, buses, etc.), especially when the radar is co-located with the BS as for ISAC,  and its height position is above the target, as in a V2I scenario. In particular, the strongest back-scattering from the VE turns out to be generated by the double bounce with the ground, whose peak in the radar image does not correspond to the antenna position on-board the VE. Based on these considerations from experiments, we model the uncertainty on the back-scattering point across the VE with an additional error on the radar-estimated antenna position. On average, this effect roughly doubles the required beam training time at the BS and leads to a $2-3$ dB beamforming gain loss in the considered simulations settings. This gets worse as the resolution of both communication and radar systems increases.
\item We propose an improved BM solution that consists of the optimal statistical combination of both radar- and position-based estimates. Remarkably, the proposed multi-technology BM guarantees the best trade off performance, providing a comparable training time and a reduced gain loss of 3 dB, on average, compared to stand-alone radar or positioning solutions. 
\end{itemize}

\subsection*{Organization}
The remainder of the paper is structured as follows. Section~\ref{sect:system_model} defines the system and channel models. The different BM methods are highlighted in Section~\ref{sect:beam_training_schemes}. The modelling of the radar back-scattering point on the VE for the purpose of BM is discussed in Section~\ref{sect:radar_backscattering} with the support of experimental campaign. Section~\ref{sect:results} present and discusses the simulation results. Finally, Section~\ref{sect:conclusion}  concludes the paper.  

\subsection*{Notation}
Bold upper and lower-case letters stand for matrices and column vectors, respectively. $\left[\mathbf{{A}}\right]_{(i,j)}$ denotes the $(i,j)$ entry of matrix $\mathbf{A}$. Matrix transposition and conjugate transposition of $\mathbf{A}$ are indicated as $\mathbf{A}^{\mathrm{T}}$ and $\mathbf{A}^{\mathrm{H}}$, respectively . $\mathbf{A}\otimes\mathbf{B}$ is the Kronecker (tensor) product between matrix $\mathbf{A}$ and matrix $\mathbf{B}$.
With $\mathbf{a}\sim\mathcal{CN}(\boldsymbol{\mu},\mathbf{C})$ we denote a multi-variate complex Gaussian random variable $\mathbf{a}$ with mean $\boldsymbol{\mu}$ and covariance $\mathbf{C}$. $\mathbf{A} \succcurlyeq \mathbf{B}$ implies that $\mathbf{A}-\mathbf{B}$ is positive semidefinite. $\mathbf{I}_N$ stands for the identity of dimensions $N$. $\mathbb{E}$ denotes the expectation operator. $\mathbb{R}$ and $\mathbb{C}$ denote, respectively, the set of real and complex numbers. $\delta_n$ is the Kronecker delta.

\section{System and channel model}\label{sect:system_model}

We consider the single-user MIMO system depicted in Fig.~\ref{fig:system_model}, representing the downlink communication between a single array panel of the BS and the VE. Each panel of the BS is  half-wavelength spaced Uniform Planar Array (UPA) with $N_{\mathrm{BS}} = N^{\mathrm{az}}_{\mathrm{BS}} \times N^{\mathrm{el}}_{\mathrm{BS}}$ antenna elements, where $N^{\mathrm{az}}_{\mathrm{BS}}$ and $N^{\mathrm{el}}_{\mathrm{BS}}$ denote the number of antennas in azimuth and elevation, respectively. Similarly, the VE is equipped with an UPA with $N_{\mathrm{VE}} = N^{\mathrm{az}}_{\mathrm{VE}} \times N^{\mathrm{el}}_{\mathrm{VE}}$ antennas. 
Let the complex transmitted symbol be $s$, characterized by a 
transmitted power $\sigma_s^2$. Symbol $s$ is beamformed towards the VE by $\mathbf{f} \in \mathbb{C}^{N_{\mathrm{BS}} \times 1}$, the transmitted signal is 
\begin{equation}
    \mathbf{x} = \mathbf{f}\,s.
\end{equation}
The propagation is over a block-faded spatially-sparse MIMO channel $\mathbf{H} \in \mathbb{C}^{N_{\mathrm{VE}} \times N_{\mathrm{BS}}}$, that at mmW/sub-THz bands can be modelled as the sum of $P$ paths as~\cite{channelMiz} 
\begin{equation}\label{eq:channel_matrix_freq}
    \mathbf{H} = \sum_{p=1}^{P}\alpha_p\,\mathbf{a}_r(\pmb{\theta}_{t,p})\mathbf{a}^{\mathrm{H}}_t(\pmb{\theta}_{r,p}),
\end{equation}
where \textit{(i)} $\alpha_p$ is the complex gain of the $p$th path, including the Doppler shift induced by the VE's motion; \textit{(ii)} $\mathbf{a}_r(\pmb{\theta}_{r,p}) \in \mathbb{C}^{N_{\mathrm{VE}}\times 1}$ and $\mathbf{a}_t(\pmb{\theta}_{t,p}) \in \mathbb{C}^{N_{\mathrm{BS}}\times 1}$ are the steering vectors for the receiver and transmitter UPA, respectively function of the Angles of Arrival (AoAs) $ \pmb{\theta}_{r,p} = [\vartheta_{r,p}, \varphi_{r,p}]^\mathrm{T}$ and of the Angles of Departure (AoDs) $\pmb{\theta}_{t,p} = [\vartheta_{t,p}, \varphi_{t,p}]^\mathrm{T}$ (for azimuth and elevation). Steering vectors are structured as
\begin{equation}
    \mathbf{a}(\pmb{\theta}) = \frac{1}{\sqrt{N_\mathrm{az}}}\mathbf{a}_{N_\mathrm{az}}(\vartheta)\otimes \frac{1}{\sqrt{N_\mathrm{el}}}\mathbf{a}_{N_\mathrm{el}}(\varphi),
\end{equation}
where $N_\mathrm{az}$ and $N_\mathrm{el}$ denote the number of azimuth and elevation antennas of either BS or VE, $\mathbf{a}_{N_\mathrm{az}}(\vartheta)$ and $\mathbf{a}_{N_\mathrm{el}}(\varphi)$ are half-wavelength spaced steering vectors toward $\vartheta$ and $\varphi$ for linear arrays of $N_\mathrm{az}$ and $N_\mathrm{el}$ elements, respectively. Notice that AoDs and AoAs are defined in two distinct \textit{local} reference systems, aligned with the BS and VE UPAs, respectively.
\begin{figure}
    \centering
    \includegraphics[width=0.85\columnwidth]{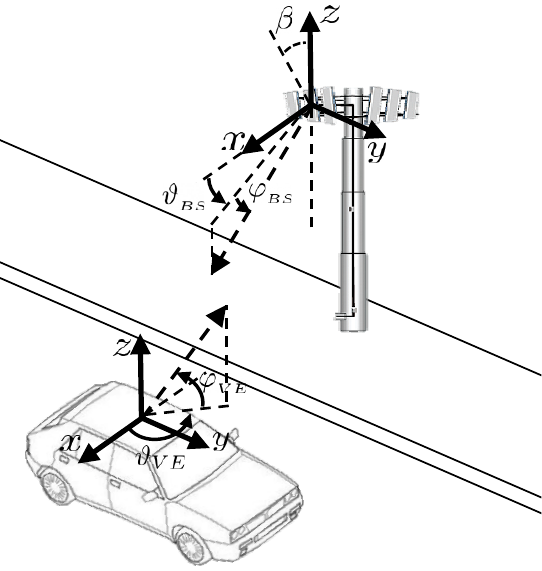}
    \caption{Definition of reference system for UPA, where the angle $\beta$ defines the mechanical tilt of the BS's panel.}
    \label{fig:sistema_rif}
\end{figure}
The received signal is
\begin{equation}\label{eq:rx_signal_freq}
    y = \mathbf{w}^\mathrm{H} \mathbf{H} \mathbf{f} \,s + \mathbf{w}^\mathrm{H} \mathbf{n},
\end{equation}
where $\mathbf{w} \in \mathbb{C}^{N_{\mathrm{VE}} \times 1}$ is the beamformer at VE, and $\mathbf{n} \sim \mathcal{CN}\left(\mathbf{0}, \sigma^2_n\mathbf{I}_{N_{\mathrm{VE}}}\right)$ is the additive white complex circular Gaussian noise, with $\sigma^2_n$ being the noise power at each VE antenna.

Both BS and VE beamformers are chosen in orthogonal codebooks $\mathcal{F}$ and $\mathcal{W}$~\cite{dftCod}, thus $\mathbf{f}\in\mathcal{F}$ and $\mathbf{w}\in\mathcal{W}$. For instance, the BS codebook is structured as
\begin{equation}\label{eq:codebook_BS}
    \mathcal{F}=\left\{\mathbf{f}_{n} = \mathbf{f}_{i}\otimes\mathbf{f}_{j}\bigg\lvert \mathbf{f}_{i} \in \mathcal{F}_{\vartheta},\, \mathbf{f}_{j} \in \mathcal{F}_{\varphi} \right\},
\end{equation}
where the two azimuth and elevation codebooks $\mathcal{F}_{\vartheta}$ and $\mathcal{F}_{\varphi}$ are 
\begin{align}
    \mathcal{F}_\vartheta & = \bigg\{\mathbf{f}_{i} \bigg\lvert \mathbf{f}^\mathrm{H}_{i}\mathbf{f}_{p}=N^\mathrm{az}_\mathrm{BS}\,\delta_{i-p},\, i,p = -\frac{N^\mathrm{az}_{\mathrm{BS}}}{2},...,\frac{N^\mathrm{az}_{\mathrm{BS}}}{2}\bigg\},\label{eq:codebook_BS_az}\\
    \mathcal{F}_\varphi & = \bigg\{\mathbf{f}_{j} \bigg\lvert \mathbf{f}^\mathrm{H}_{j}\mathbf{f}_{\ell}=N^\mathrm{el}_\mathrm{BS}\,\delta_{j-\ell},\,j,\ell = -\frac{N^\mathrm{el}_{\mathrm{BS}}}{2},...,\frac{N^\mathrm{el}_{\mathrm{BS}}}{2}\bigg\}\label{eq:codebook_BS_el}.
\end{align}
Vectors $\mathbf{f}_i$ and $\mathbf{f}_j$ in~\eqref{eq:codebook_BS_az} and~\eqref{eq:codebook_BS_el} are drawn from Discrete Fourier Transform (DFT) matrices of size $N^\mathrm{az}_\mathrm{BS}$ and $N^\mathrm{el}_\mathrm{BS}$, respectively, thus $\mathcal{F}_\vartheta$ and $\mathcal{F}_\varphi$ span the following sets of azimuth and elevation angles:
\begin{align}
    \Theta_{\mathrm{BS}} &= \left\{\vartheta_i \hspace{-0.1cm}=\hspace{-0.1cm} \arcsin\left(i\frac{2}{N^\mathrm{az}_{\mathrm{BS}}}\right)\bigg\lvert \,\,i = -\frac{N^\mathrm{az}_{\mathrm{BS}}}{2},...,\frac{N^\mathrm{az}_{\mathrm{BS}}}{2}\right\} \label{eq:BS_angles_az},\\
    \Phi_{\mathrm{BS}} &= \hspace{-0.1cm}\left\{\varphi_j \hspace{-0.1cm}=\hspace{-0.1cm} \arcsin\left(j\frac{2}{N^\mathrm{el}_{\mathrm{BS}}}\right)\bigg\lvert \,\,j = -\frac{N^\mathrm{el}_{\mathrm{BS}}}{2},...,\frac{N^\mathrm{el}_{\mathrm{BS}}}{2}\right\}.\label{eq:BS_angles_el}
\end{align}
Thus, $\mathbf{f}_i=\mathbf{a}_{N^\mathrm{az}_{\mathrm{BS}}}(\vartheta_i)$, $\mathbf{f}_j=\mathbf{a}_{N^\mathrm{el}_{\mathrm{BS}}}(\varphi_j)$.
The beamforming codebook used by the VEs $\mathcal{W}$ is similarly structured based on VE's array configuration. Similar angular codebooks $\Theta_{\mathrm{VE}}$ and $\Phi_{\mathrm{VE}}$ can be defined for the VE starting from $\mathcal{W}$. Figure~\ref{fig:sistema_rif} shows the BS and VE local reference systems.

%

\section{Beam Management Schemes}\label{sect:beam_training_schemes}


In this section, we address the problem of V2I BM in a vehicular scenario. To this aim, let us consider a VE, moving within the BS coverage area, that performs a BM procedure in case of IA or whenever the link quality degrades under a predefined threshold, as explained in Section~\ref{sect:results}. 
At each BM instant, the optimal beamformers at both BS and VE side to be used in~\eqref{eq:rx_signal_freq} are the maximizers of the received power at the VE:
\begin{equation} \label{eq:optimal_beamformers}
    (\mathbf{f}_{opt},\mathbf{w}_{opt})=\underset{\substack{\mathbf{f}\in\mathcal{F}\\\mathbf{w}\in\mathcal{W}}}{\mathrm{argmax}}\,\big\lvert \mathbf{w}^{\mathrm{H}}\,\mathbf{H}\,\mathbf{f}s+\mathbf{w}^\mathrm{H} \mathbf{n}\big\rvert^2,
\end{equation}
searched within codebooks $\mathcal{F}$ and $\mathcal{W}$. The optimal SNR is therefore
\begin{equation}
    \gamma(\mathbf{f}_{opt},\mathbf{w}_{opt}) = \frac{\lvert\mathbf{w}_{opt}^{\mathrm{H}}\,\mathbf{H}\,\mathbf{f}_{opt}\rvert^2 \sigma^2_s}{N_\mathrm{VE}\,\sigma^2_n }.
\end{equation}
The cost of performing a search across codebooks $\mathcal{F}$ and $\mathcal{W}$ can be expressed in terms of beam training time $T_{\mathrm{train}}$, defined as
\begin{equation} \label{eq:TrTime}
    T_{\mathrm{train}} = T \, \lvert\mathcal{F}\rvert \lvert\mathcal{W}\rvert,
\end{equation}
where $T$ is the time taken for testing a single BS-VE beam pair and $\lvert\cdot\rvert$ is the cardinality of the given set. The assumption in~\eqref{eq:optimal_beamformers} is that $(\mathbf{f}_{opt},\mathbf{w}_{opt})$ do not change within $T_{\mathrm{train}}$. Considering the typical parameters of V2I communication, it is immediate to see that $\lvert\mathcal{F}\rvert \lvert\mathcal{W}\rvert$ shall be bounded. For instance, 3GPP Rel. 17 considers $T$ equal to the duration of a single Orthogonal Frequency Division Multiplexing (OFDM) slot (e.g., $62.5$ $\mu$s for a subcarrier spacing  of $\Delta f =240$ kHz). In this setting, for $N^\mathrm{az}_{\mathrm{BS}}=16$, a VE crossing the BS boresight at $10$ m distance and $50$ km/h speed is served by the same beam for a spatial interval of $\approx 1.2$ m, thus, for $72$ ms. With $T=62.5$ $\mu$s, it implies $\lvert\mathcal{F}\rvert \lvert\mathcal{W}\rvert \leq 1100$, e.g., $\lvert\mathcal{F}\rvert=64$, $\lvert\mathcal{W}\rvert=16$. Note that, in the latter settings, the spectral efficiency of the V2I system rapidly drops due to an excessively long BM procedure~\cite{Heath2020leveragingsensing}.

The goal of the BM is to reduce the training time $T_{\mathrm{train}}$ by finding suitable \textit{instantaneous} subsets of $\mathcal{F}$ and $\mathcal{W}$, namely
\begin{equation}
    \mathcal{F}_{\mathrm{sub}}\subseteq\mathcal{F},\,\,\,\,\mathcal{W}_{\mathrm{sub}}\subseteq\mathcal{W},
\end{equation}
such that, at each training time, $\lvert\mathcal{F}_{\mathrm{sub}}\rvert\ll\lvert\mathcal{F}\rvert$, $\lvert\mathcal{W}_{\mathrm{sub}}\rvert\ll\lvert\mathcal{W}\rvert$. The beamformers can be estimated by searching in $\mathcal{F}_{\mathrm{sub}}$ and $\mathcal{W}_{\mathrm{sub}}$ as
\begin{equation} \label{eq:selected_beamformers}
    (\hat{\mathbf{f}},\hat{\mathbf{w}})=\underset{\substack{\mathbf{f}\in\mathcal{F}_\mathrm{sub}\\\mathbf{w}\in\mathcal{W}_\mathrm{sub}}}{\mathrm{argmax}}\,\big\lvert \mathbf{w}^{\mathrm{H}}\,\mathbf{H}\,\mathbf{f}s+\mathbf{w}^\mathrm{H} \mathbf{n}\big\rvert^2.
\end{equation}
%
%
%
The corresponding beamforming gain loss is defined by the mismatch in the chosen beamformers as
\begin{equation} \label{eq:gain_loss}
    \Delta \mathrm{G} = \frac{\lvert\mathbf{w}^\mathrm{H}_{opt}\mathbf{H} \,\mathbf{f}_{opt}\rvert^2}{\lvert\hat{\mathbf{w}}^\mathrm{H}\,\mathbf{H} \,\hat{\mathbf{f}}\rvert^2},
\end{equation}
which is $\Delta G=1$ when $\hat{\mathbf{f}}=\mathbf{f}_{opt}$ and $\hat{\mathbf{w}}=\mathbf{w}_{opt}$.
Notice that the BM in~\eqref{eq:selected_beamformers} is triggered whenever the instantaneous SNR falls below a pre-defined value with respect to the SNR at the previous training instant. 

In the following, we detail different blind and sensor-assisted beam methods to select $\mathcal{F}_\mathrm{sub}$, assuming the VE perfectly aligned with the BS, i.e., $\mathbf{w}=\mathbf{w}_{opt}$ in~\eqref{eq:rx_signal_freq}. The procedures aim to select the angular codebooks at the BS, namely
\begin{align}
    &\Theta^{\mathrm{sub}}_{\mathrm{BS}}\subseteq\Theta_{\mathrm{BS}}, \,\,\,\, \Phi^\mathrm{sub}_{\mathrm{BS}}\subseteq\Phi_{\mathrm{BS}},
\end{align}
that are related to beam codebooks by~\eqref{eq:codebook_BS_az}-\eqref{eq:codebook_BS_el} and~\eqref{eq:BS_angles_az}-\eqref{eq:BS_angles_el}.

\subsection{Blind methods}





Blind beam management approaches, currently defined in the 3GPP standard~\cite{TS_38213}, do not rely on any type of side information. These are conceptually divided into Exhaustive Search (ES) approaches, where a full beam sweeping is performed spanning the whole codebook $\mathcal{F}$ and Gradient Search (GS) approaches, where the beam sweeping at the BS covers suitable subsets of the full codebooks.

\textbf{Exhaustive Search (ES)}: is the most simple and intuitive method and it is adopted for the 3GPP standard IA procedure~\cite{TS_38213}. 
The beamformer $\hat{\mathbf{f}}$ in~\eqref{eq:selected_beamformers}
is searched across the full BS codebook, thus 
\begin{equation}
\begin{split}
    & \Theta^\mathrm{sub}_{\mathrm{BS}} = \Theta_{\mathrm{BS}} ,\,\,\,\,  \Phi^\mathrm{sub}_{\mathrm{BS}} = \Phi_{\mathrm{BS}}.
\end{split}
\end{equation}

\textbf{Gradient Search (GS)}: used in beam tracking mode in the 5G NR standard~\cite{TS_38213}. 
The GS is initialized with an ES during the IA to find the global optimum.
The strategy behind the GS approach is to focus the search around the azimuth and elevation angles, $\hat{\vartheta}$ and $\hat{\varphi}$, determined at the previous beam training instant. The subset of BS angular codebooks is computed as
\begin{align} \label{eq:codebookGS}
    \Theta^\mathrm{sub}_\mathrm{BS} &= \left\{\hat{\vartheta} + \vartheta_k \bigg\lvert \,\,k = -\frac{K}{2}, \dots \frac{K}{2}\right\},\\
    \Phi^\mathrm{sub}_\mathrm{BS} &= \left\{\hat{\varphi} + \varphi_{k} \bigg\lvert \,\,k = -\frac{K}{2}, \dots \frac{K}{2}\right\},
\end{align}
where $\hat{\vartheta} +\vartheta_k\in\Theta_\mathrm{BS}$, $\hat{\varphi} +\varphi_k\in\Phi_\mathrm{BS}$ and $K^2$ represents the GS codebook cardinality. High values of $K$ lead to higher complexity in~\eqref{eq:optimal_beamformers}, while low values can lead to losing the global optimum and thus error propagation. 
\begin{figure}[!t] 
\centering
\includegraphics[width=0.7\columnwidth]{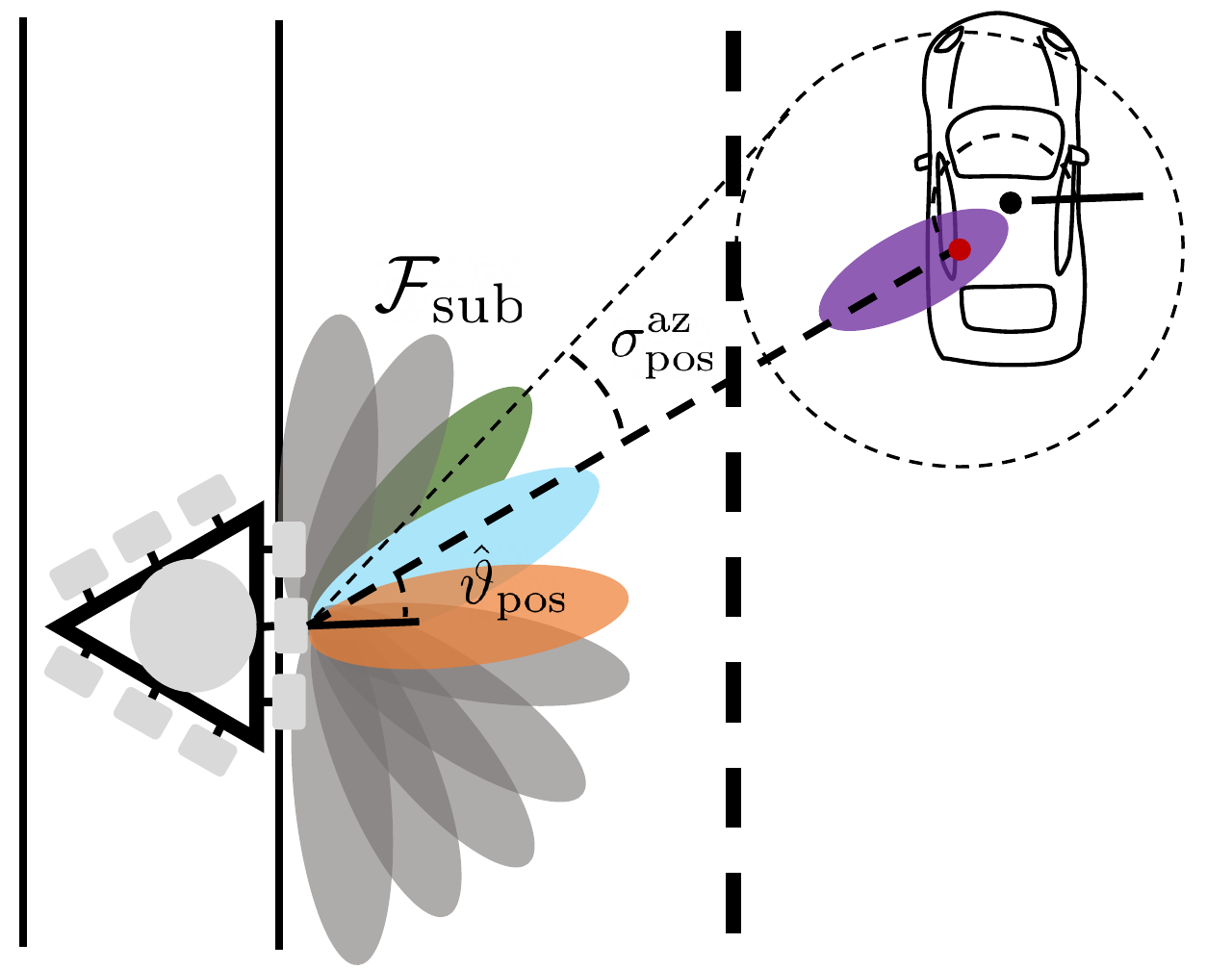}
\caption{Position-assisted BM method: the reduced BS codebook depends on localization accuracy $\sigma^{az}_{\mathrm{pos}}$.}
\label{fig:pos}
\end{figure}

\subsection{Position-assisted methods}

Position-assisted BM methods relay on the side information from positioning systems on board the VE. In addition to GNSS, other sensors can be exploited to enhance the accuracy and robustness of the VE's estimated position, such as inertial sensors, wheel odometer, steering angle sensor, camera, lidar, automotive radars etc.~\cite{Brambilla2020,Gao2022_positioning}. Based on the periodically estimated VE position, the BS codebook is dynamically updated. 

Periodically, at rate $R_{\mathrm{pos}}$, the VE obtains an estimate of its ego position in global Cartesian coordinates, modelled as
\begin{equation} \label{eq:est_pos}
    \hat{\mathbf{p}}_{\mathrm{pos}}=\begin{bmatrix}\hat{x}_{\mathrm{pos}}\\\hat{y}_{\mathrm{pos}}\\\hat{z}_{\mathrm{pos}}\end{bmatrix}=\mathbf{p} + \mathbf{e}_{\mathrm{pos}}, 
\end{equation}
where $\mathbf{p}=[x,y,z]^\mathrm{T}$ is the true position of the VE's mmW/sub-THz antenna, affected by $\mathbf{e}_{\mathrm{pos}}\sim \mathcal{N}(\mathbf{0}, \mathbf{C}_{\mathrm{pos}})$ as measurement error. For the purpose of this work, the covariance matrix $\mathbf{C}_{\mathrm{pos}}\in \mathbb{R}^{3\times3}$ is approximated as 
\begin{equation}\label{eq:GPS_cov}
    \mathbf{C}_{\mathrm{pos}} \approx \text{diag}\left([\sigma_{x}^{2},\sigma_{y}^{2},\sigma_{z}^{2}]\right)
\end{equation}
where $\sigma_{x}$, $\sigma_{y}$ and $\sigma_{z}$ model position uncertainties along the $x$, $y$ and $z$  directions, respectively. 
The estimated position in~\eqref{eq:est_pos} is mapped from global Cartesian coordinates into the local BS spherical ones knowing the BS position and the possible (fixed) rotation between global and BS Cartesian axis, obtaining
\begin{equation}\label{eq:est_pos_spherical}
    \hat{\mathbf{u}}_{\mathrm{pos}}= \begin{bmatrix}\hat{\rho}_{\mathrm{pos}}\\ \hat{\vartheta}_{\mathrm{pos}}\\\hat{\varphi}_{\mathrm{pos}}\end{bmatrix}=\mathbf{u} + \mathbf{e}^u_{\mathrm{pos}}, 
\end{equation}
where $\mathbf{u} = [\rho,\vartheta,\varphi]^\mathrm{T}$ denotes the true position of the mmW/sub-THz VE antenna in BS-centered spherical coordinates. The error $\mathbf{e}^u_{\mathrm{pos}}$ is characterized by the following covariance matrix
\begin{equation}\label{eq:GPS_cov_spherical}
    \mathbf{C}^u_{\mathrm{pos}} = \mathbf{T}\mathbf{R} \,\mathbf{C}_{\mathrm{pos}}\mathbf{R}^\mathrm{T}\mathbf{T}^\mathrm{T},
\end{equation}
where $\mathbf{R}\in\mathbb{R}^{3\times 3}$ is the rotation matrix from global Cartesian coordinates to local (BS) Cartesian ones, $\mathbf{T}\in\mathbb{R}^{3\times 3}$ is the Jacobian from Cartesian to spherical coordinates. Diagonal elements $[\mathbf{C}^u_{\mathrm{pos}}]_{(2,2)} = \left( \sigma^\mathrm{az}_{\mathrm{pos}}\right)^2$ and $[\mathbf{C}^u_{\mathrm{pos}}]_{(3,3)} = \left( \sigma^\mathrm{el}_{\mathrm{pos}}\right)^2$
define the angular uncertainties in the UE's antenna position (azimuth and elevation), used to retrieve the effective codebooks' sizes.
When $\sigma_{x}=\sigma_{y}=\sigma$ in \eqref{eq:GPS_cov}, the positioning error can be split into horizontal and vertical as $\sigma_h = \sqrt{2}\sigma$ and $\sigma_v = \sigma_z$. In this case, the angular positioning errors simplify as
\begin{align} 
    \sigma^\mathrm{az}_{\mathrm{pos}} & = \arctan\left(\frac{\sigma_{h}}{\sqrt{\hat{x}_\mathrm{pos}^{2}+\hat{y}_\mathrm{pos}^{2}}}\right) \label{eq:sigma:pos_az},\\
    \sigma^\mathrm{el} _{\mathrm{pos}} & = \arctan\left(\frac{\sigma_{v}}{\hat{z}_{\mathrm{pos}}}\right)\label{eq:sigma:pos_el},
\end{align}
illustrated in Fig.~\ref{fig:pos}.

The BS obtains the instantaneous angular codebooks as 
\begin{equation}\label{eq:BS_codebook_pos}
\begin{split}
    \Theta^\mathrm{sub}_{\mathrm{BS}} = \Theta_{\mathrm{BS}}\, \cap \,\left[\hat{\vartheta}_{\mathrm{pos}}-\sigma^\mathrm{az}_{\mathrm{pos}},\hat{\vartheta}_{\mathrm{pos}}+\sigma^\mathrm{az}_{\mathrm{pos}}\right],\\
    \Phi^\mathrm{sub}_{\mathrm{BS}} = \Phi_{\mathrm{BS}}\, \cap \,\left[\hat{\varphi}_{\mathrm{pos}}-\sigma^\mathrm{el}_{\mathrm{pos}},\hat{\varphi}_{\mathrm{pos}}+\sigma^\mathrm{el}_{\mathrm{pos}}\right],
\end{split}
\end{equation}
i.e., testing the beams whose spanned angles are within the angular uncertainty region provided by the positioning system.
%
    %
    %
    %

%
\subsection{Radar-assisted methods}


Another solution to provide vehicle state-awareness at the BS is to use a co-located MIMO radar. Differently from position-assisted BM methods leveraging on VE sensors, radar at BS approaches do not need any control signaling between the VE and BS, thus no control overhead. For each target within the field-of-view, radars provide an estimation of the range, angular position, and radial velocity.  Although velocity estimation (from Doppler spectrum analysis) could enable target tracking over time, this is beyond the scope of this work only to provide the fairest possible comparison between conventional blind BM methods and sensor-assisted ones. Therefore, in the following, we model the range-angle estimation from a MIMO radar and the corresponding usage for BM.

Let us therefore assume a MIMO radar at the BS, oriented as the BS panel, and equipped with $N_{\mathrm{rad}} = N^\mathrm{az}_{\mathrm{rad}} \times N^\mathrm{el}_{\mathrm{rad}}$ real/virtual channels, employing a bandwidth $B_{\mathrm{rad}}$. 
Let us assume that the VE is the only target in the environment, and that the MIMO radar is able to detect the position of the VE mmW/sub-THz antenna. The estimation of the VE antenna position, obtained at rate $R_{\mathrm{rad}}$, can be modelled in local (BS/radar) spherical coordinates as
\begin{equation}\label{eq:radar_estimated_pos}
    \hat{\mathbf{u}}_{\mathrm{rad}} = \begin{bmatrix}\hat{\rho}_\mathrm{rad}\\
    \hat{\vartheta}_\mathrm{rad}\\
    \hat{\varphi}_\mathrm{rad}\end{bmatrix} = \mathbf{u} + \mathbf{e}^u_{\mathrm{res}},
\end{equation}
where $\mathbf{e}^u_{\mathrm{res}}\sim\mathcal{N}(\mathbf{0},\mathbf{C}^u_{\mathrm{res}})$ is the Gaussian zero-mean position error that models the uncertainty on the position of the VE antenna, with covariance matrix
%
\begin{equation}\label{eq:rad_error_cov}
    \mathbf{C}^{u}_{\mathrm{res}} = \text{diag}\left( \left[\left( \sigma^\rho_{\mathrm{res}}\right)^2, \left( \sigma^\mathrm{az}_{\mathrm{rad}}\right)^2, \left( \sigma^\mathrm{el}_{\mathrm{res}}\right)^2\right]\right).
\end{equation}
In~\eqref{eq:rad_error_cov},  
\begin{equation}\label{eq:range_resolution}
    \sigma^\rho_{\mathrm{res}} = \frac{c}{2 B_\mathrm{rad}}
\end{equation}
is the range resolution, ruled by bandwidth $B_\mathrm{rad}$, $c$=$3 \cdot 10^8$ m/s, while $\sigma_{\mathrm{res}}^{\mathrm{az}}$, and $\sigma_{\mathrm{res}}^{\mathrm{el}}$ are the angular resolutions along azimuth and elevation, respectively
\begin{align}\label{eq:radar_ang_unce}
    \sigma^\mathrm{az}_{\mathrm{res}} & = \frac{\lambda}{2 N^{\mathrm{az}}_{\mathrm{rad}}\Delta \cos\vartheta},\\
  \sigma^\mathrm{el}_{\mathrm{res}} & = \frac{\lambda}{2 N^{\mathrm{el}}_{\mathrm{rad}}\Delta \cos\varphi},
\end{align}
where $\Delta$ is the inter-channel distance (either the physical spacing between adjacent antennas or the equivalent spacing between two virtual channels). But on~\eqref{eq:radar_estimated_pos} and~\eqref{eq:rad_error_cov} we are implicitly assuming that \textit{(i)} the back-scattered signal received by the radar is originated \textit{only} by the VE's antenna and \textit{(ii)} the radar is always capable of detecting the target (i.e. the VE's antenna) with the sensing SNR sufficiently high. For MIMO radars, the latter condition \textit{(ii)} can be achieved by incoherently averaging multiple range-angle images obtained by a 2D-DFT over fast-time and angles, respectively~\cite{tagliaferri_navigation-aided_2021,manzoni2022motion}, implying that the uncertainty on antenna's position is dominated by the resolution. Notice that the angular resolution degrades for off-boresight azimuth and elevation pointing angles.

Similarly to~\eqref{eq:BS_codebook_pos}, the BS codebook is retrieved as:
\begin{equation}\label{eq:BS_codebook_rad}
\begin{split}
    \Theta^\mathrm{sub}_{\mathrm{BS}} = \Theta_{\mathrm{BS}}\, \cap \,\left[\hat{\vartheta}_{\mathrm{rad}}-\sigma^\mathrm{az}_{\mathrm{res}},\hat{\vartheta}_{\mathrm{rad}}+\sigma^\mathrm{az}_{\mathrm{res}}\right],\\
    \Phi^\mathrm{sub}_{\mathrm{BS}} = \Phi_{\mathrm{BS}}\, \cap \,\left[\hat{\varphi}_{\mathrm{rad}}-\sigma^\mathrm{el}_{\mathrm{res}},\hat{\varphi}_{\mathrm{rad}}+\sigma^\mathrm{el}_{\mathrm{res}}\right].
\end{split}
\end{equation}

\subsection{Multi-Technology (MT) method} \label{sec:MT}

In addition to position- and radar-assisted BM solutions, we propose a possible improvement that leverages both radar and positioning systems. The optimal statistical combination of the dual technologies VE's estimated positions is
\begin{equation}
    \hat{\mathbf{u}}_\mathrm{MT} =\begin{bmatrix}\hat{\rho}_{\mathrm{MT}}\\ \hat{\vartheta}_{\mathrm{MT}}\\\hat{\varphi}_{\mathrm{MT}}\end{bmatrix}= \mathbf{K}_\mathrm{pos}\,\hat{\mathbf{u}}_\mathrm{pos} + \mathbf{K}_\mathrm{rad}\,\hat{\mathbf{u}}_\mathrm{rad}
\end{equation}
where 
\begin{align}
    &\mathbf{K}_\mathrm{pos} = \omega\, \left(\mathbf{C}^u_\mathrm{MT} \mathbf{C}^u_\mathrm{pos} \right)^{-1}\\
    &\mathbf{K}_\mathrm{rad}=(1-\omega)\, \left(\mathbf{C}^u_\mathrm{MT} \mathbf{C}^u_\mathrm{rad}\right)^{-1}
\end{align}
are the two fusion matrices reporting the consistent covariance
\begin{equation} \label{eq:mt_variance}
    \mathbf{C}^u_\mathrm{MT} =  \left( \omega\left(\mathbf{C}^u_\mathrm{pos}\right)^{-1} + (1-\omega)\left(\mathbf{C}^u_\mathrm{rad}\right)^{-1}\right)^{-1}, 
\end{equation}
i.e., $\mathbf{C}^u_\mathrm{MT} \succcurlyeq \mathbb{E}[\hat{\mathbf{u}}_\mathrm{MT} \hat{\mathbf{u}}^\mathrm{T}_\mathrm{MT}]$ (for optimal combination of two estimators, see e.g.~\cite{NOACK201735}). Parameter $\omega$ is a degree of freedom in the fusion design, herein we set $\omega=0.5$. But a possible improvement could be to set $\omega$ as function of the information freshness.

The BS codebook is now:
\begin{equation}\label{eq:BS_codebook_MT}
\begin{split}
    \Theta^\mathrm{sub}_{\mathrm{BS}} = \Theta_{\mathrm{BS}}\, \cap \,\left[\hat{\vartheta}_{\mathrm{MT}}-\sigma^\mathrm{az}_{\mathrm{MT}},\hat{\vartheta}_{\mathrm{MT}}+\sigma^\mathrm{az}_{\mathrm{MT}}\right],\\
    \Phi^\mathrm{sub}_{\mathrm{BS}} = \Phi_{\mathrm{BS}}\, \cap \,\left[\hat{\varphi}_{\mathrm{MT}}-\sigma^\mathrm{el}_{\mathrm{MT}},\hat{\varphi}_{\mathrm{MT}}+\sigma^\mathrm{el}_{\mathrm{MT}}\right],
\end{split}
\end{equation}
where $\left(\sigma^\mathrm{az}_{\mathrm{MT}}\right)^2 = [\mathbf{C}^u_{\mathrm{MT}}]_{(2,2)}$ and $\left(\sigma^\mathrm{el}_{\mathrm{MT}}\right)^2 = [\mathbf{C}^u_{\mathrm{MT}}]_{(3,3)}$.
The proposed Multi-Technology (MT) BM approach provides the best trade-off between training time and beamforming gain loss, as shown in Section~\ref{sect:results}.

\section{Modeling the experimented radar back-scattering of VE}\label{sect:radar_backscattering}

As mentioned in Section~\ref{sec:intro}, even in a single target scenario, it is unlikely that the scattering point on the target/VE matches the true location of the antenna $\mathbf{u}$. More realistically, the impinging wave will be back-scattered by multiple points (a continuous region) over the shape of the vehicle~\cite{berthold2017scatter}, depending on the azimuth-elevation incidence direction. Typically, when a vehicle is illuminated from the frontal direction, the back-scattering is dominated by the equivalent dihedral between the bumper and the ground or, to a lesser extent, by the reflection from the hood. 
A similar consideration applies for rear illumination. However, if the antenna is mounted on the VE's rooftop, as indicated in~\cite{14rel}, the impinging sensing wave tends to be specularly reflected by the metallic rooftop and the antenna contribution in the radar image may not be detectable. This means that, even in the ideal case where the angular resolution is infinite (i.e., $N_{\mathrm{rad}}\rightarrow\infty$, $N_{\mathrm{BS}}\rightarrow\infty$) the identification of the strongest scattering point in the range-angle map does not provide the correct pointing direction.

\subsection{Radar measurements of a moving VE}

\begin{figure}[!t]
    \centering
    \subfloat[][]{\includegraphics[width = 0.8 \columnwidth]{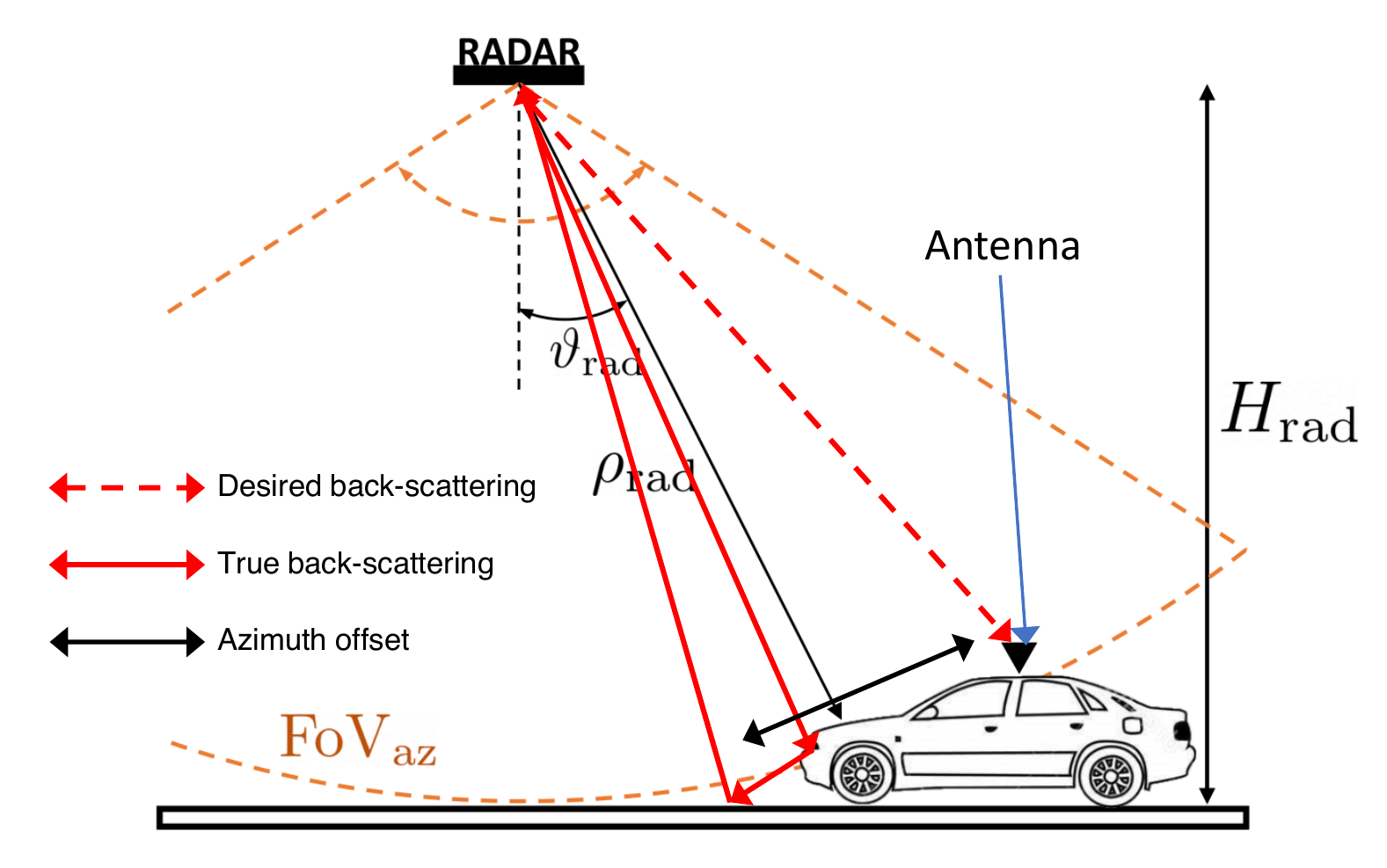}\label{subfig:radar_scenario_az}}\\
    \subfloat[][]{\includegraphics[width = 0.8 \columnwidth]{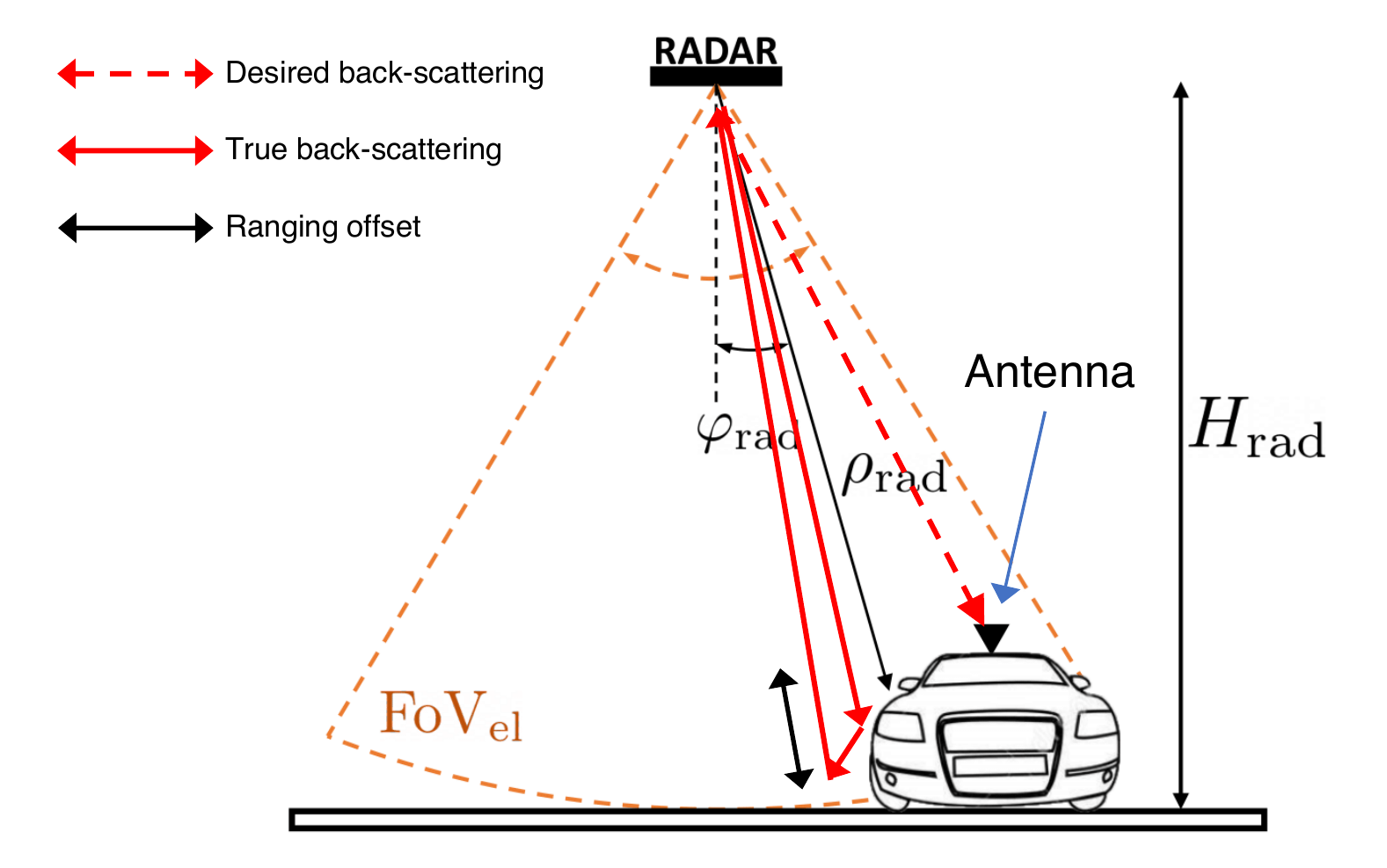}\label{subfig:radar_scenario_el}}
    \caption{Geometry of the radar acquisition: azimuth plane (\ref{subfig:radar_scenario_az}) and elevation plane (\ref{subfig:radar_scenario_el})}
    \label{fig:radar_scenario}
\end{figure}
\begin{figure}[tpb]
\centering
  \includegraphics[width=\linewidth]{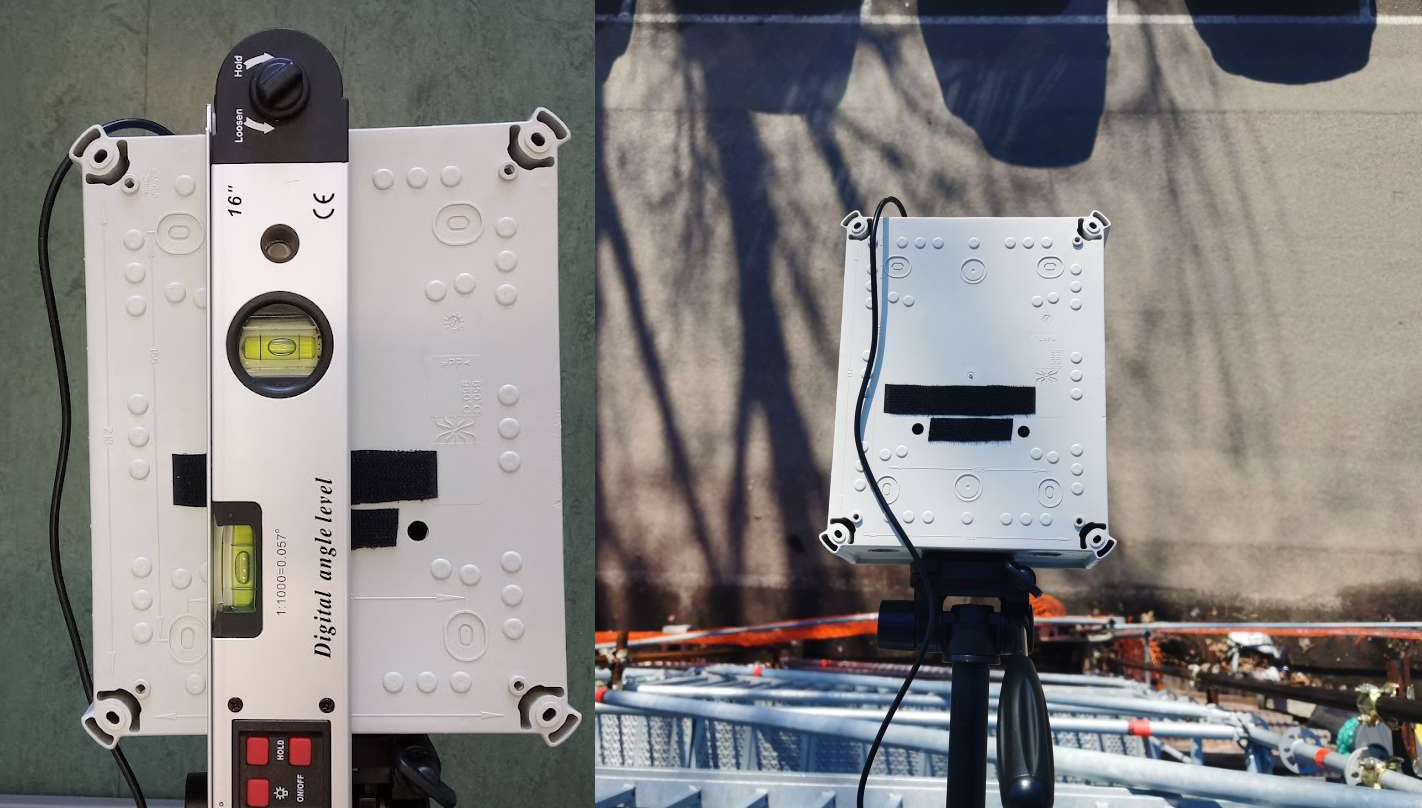}
   \caption{Photos of the measurement setup. A level is used to align the plane of measure, so that it is perpendicular to the ground plane}\label{fig:photo_setup}
   \vspace{-1pt}
\end{figure}
\begin{figure}[!t] 
\centering
\subfloat[][]{\includegraphics[width=\columnwidth]{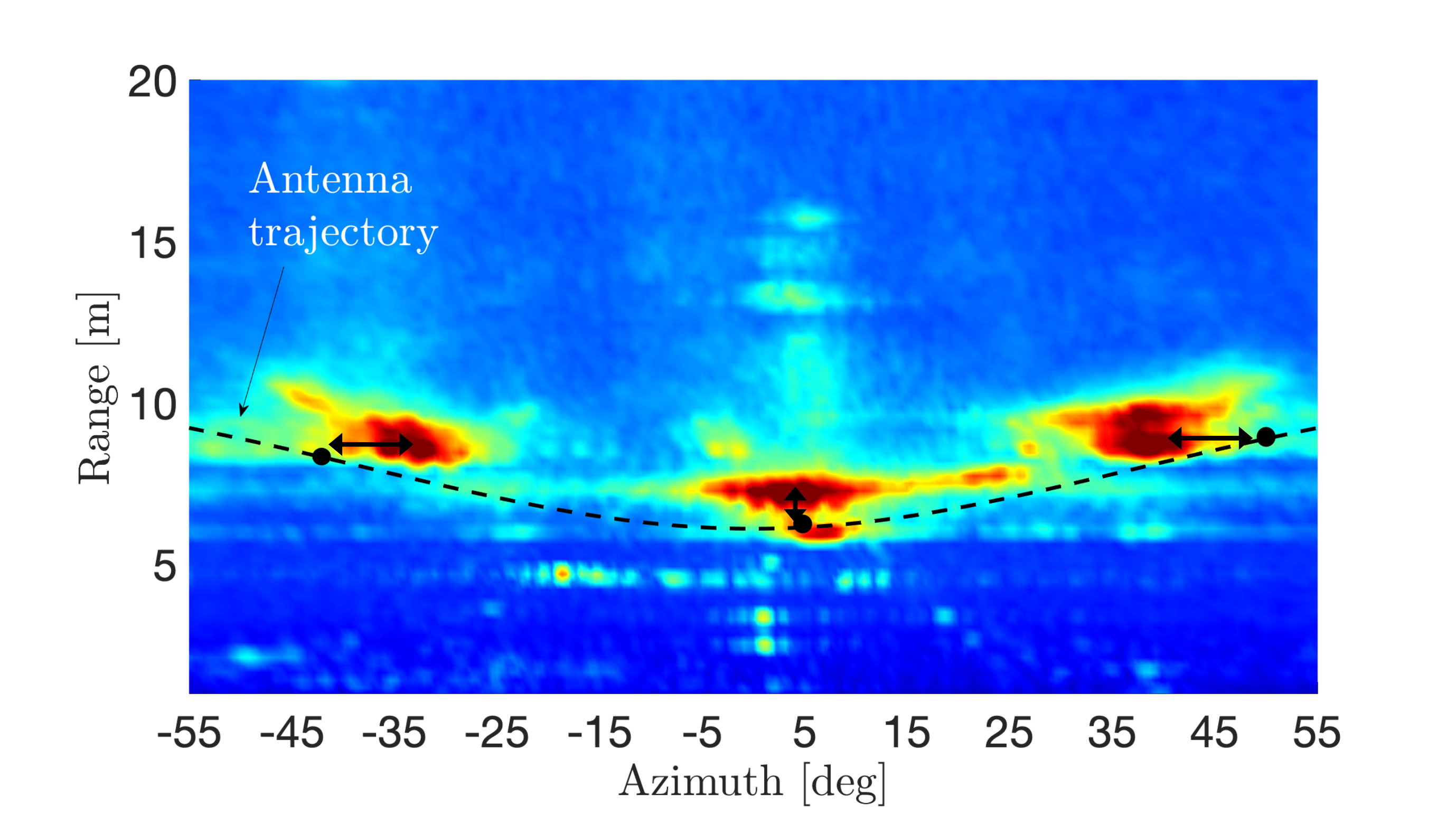}\label{subfig:radar_az_range}}\\
\subfloat[][]{\includegraphics[width=\columnwidth]{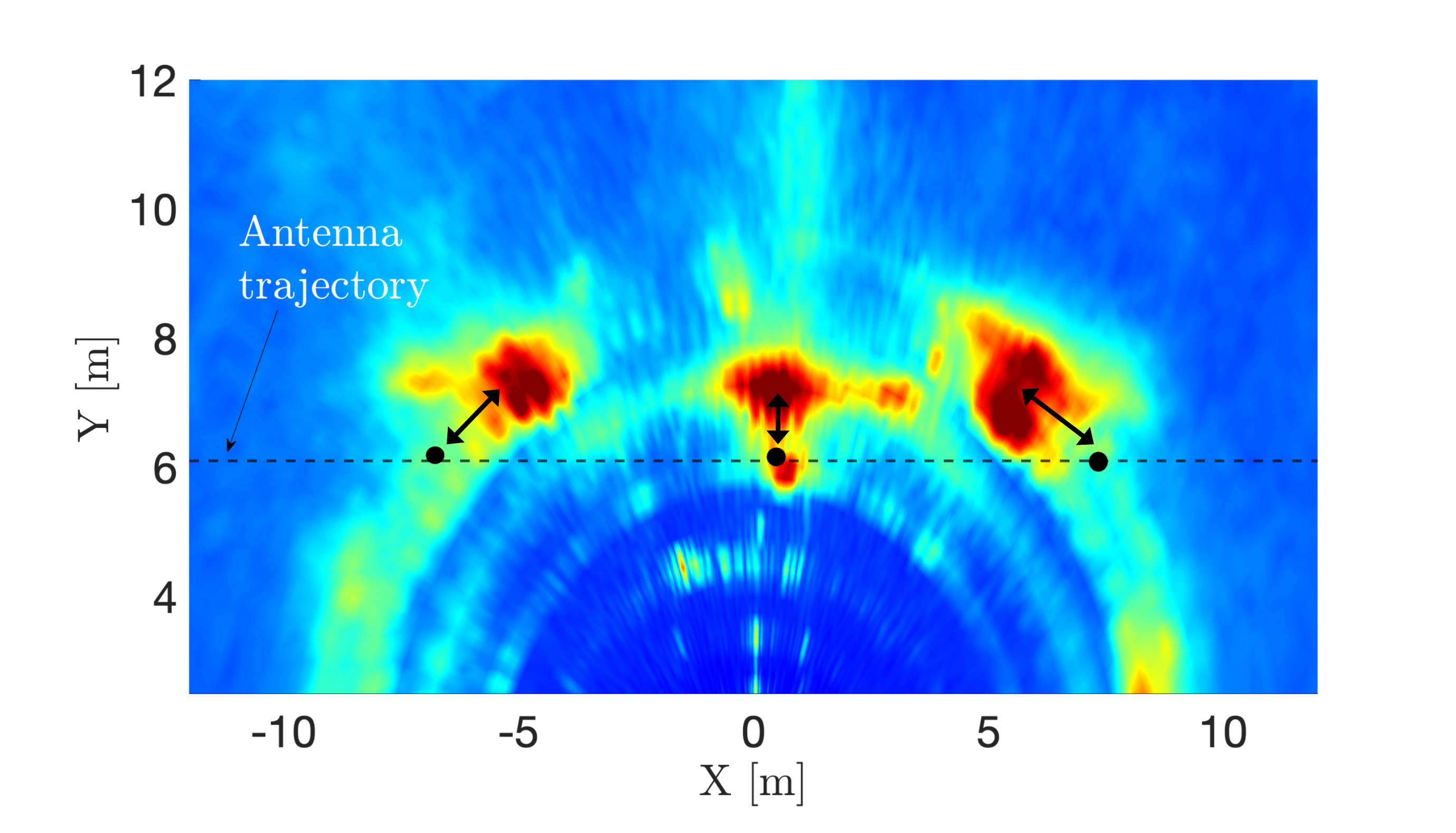}\label{subfig:radar_cartesian}}
\caption{Different radar acquisition frames of the vehicle along a straight trajectory (dashed black line) in (\ref{subfig:radar_az_range}) range-azimuth coordinates (\ref{subfig:radar_cartesian}) Cartesian coordinates. The black dots indicate the true antenna position.}
\label{fig:radar_measurement}
\end{figure}

To quantify the aforementioned phenomenon, we carried out a dedicated extensive experimental campaign aimed at acquiring radar data over different scenarios reproducing a BS-mounted radar. Herein, we report an extract of the mentioned campaign, whose scenario is illustrated in Fig.~\ref{fig:radar_scenario}. A car is travelling on a straight path in front of a MIMO radar placed at $H_{\mathrm{rad}}=7.3$~m from ground, pointing downward. In Fig.~\ref{fig:radar_scenario}, we indicate a possible antenna mounting position the center of the car's rooftop, at $H_{\mathrm{VE}}=1.5$ m from ground (but same considerations hold true for any other position such that $H_{\mathrm{VE}} < H_{\mathrm{rad}}$). The radar-VE distance on the ground is $2.5$ m. The physical setup is shown in Fig.~\ref{fig:photo_setup}. A stand and a digital goniometer to accurately measure the inclination of the radar with respect to the ground plane, whereas a laser distance meter is used to measure its height from the ground plane. The hardware set-up used for collecting MIMO radar data is the Texas Instruments' MMWCAS-RF-EVM~\cite{TI_ref_MMWCAS}, which is used in tandem with its digital signal processing board. The board contains four AWR2243 radar transceivers that are configured with $16$ receiver and $12$ transmitter antennas in total, allowing for an azimuth resolution of $1.4$ deg~\cite{TI_ref_MMWCAS}.
The azimuth field of view is $120$ deg, while the elevation one is $40$ deg. The experimental results were obtained with a bandwidth of $639.6$ MHz at the center frequency of $77.3$ GHz~\cite{TI_ref_Chirp}.

Figure~\ref{fig:radar_measurement} reports the radar map of the vehicle in both range-angle  and Cartesian coordinates, where three different positions merged in one unique radar image. As stated before, according to the true relative position of the vehicle with respect to the radar, the strongest scattering point does not coincide with the exact location of the antenna, due to double reflections with the ground. Figure~\ref{fig:radar_scenario} graphically explains this behavior. When the vehicle is off-boresight, namely the radar signal is impinging either on the frontal-side part or on the rear-side part of the vehicle, the detected back-scattering spot is subject to an azimuth shift with respect to the true antenna position (see Fig.~\ref{subfig:radar_scenario_az}). Differently, when the radar is illuminating the car's side, there is a ranging error due to the different height of the VE's antenna with respect to the ground, as illustrated in Fig.~\ref{subfig:radar_scenario_el}. In this latter case, we can notice a further detected spot due to the corner boundary between the car door and the rooftop. Moreover, notice that the size of the detected spot (few m$^2$) in the radar image tends to be much larger than the resolution cell, which is in the order of $25 \times 25$ cm$^2$ in the current settings.
The range-azimuth shift and the size of the strongest detected spot in the radar image represent, respectively, an unwanted bias and an increased uncertainty in the estimated VE's antenna location, which shall be taken into account in model of radar-estimated antenna position, as discussed in the following.

\begin{figure}[!t] 
\centering
\includegraphics[width=0.7\columnwidth]{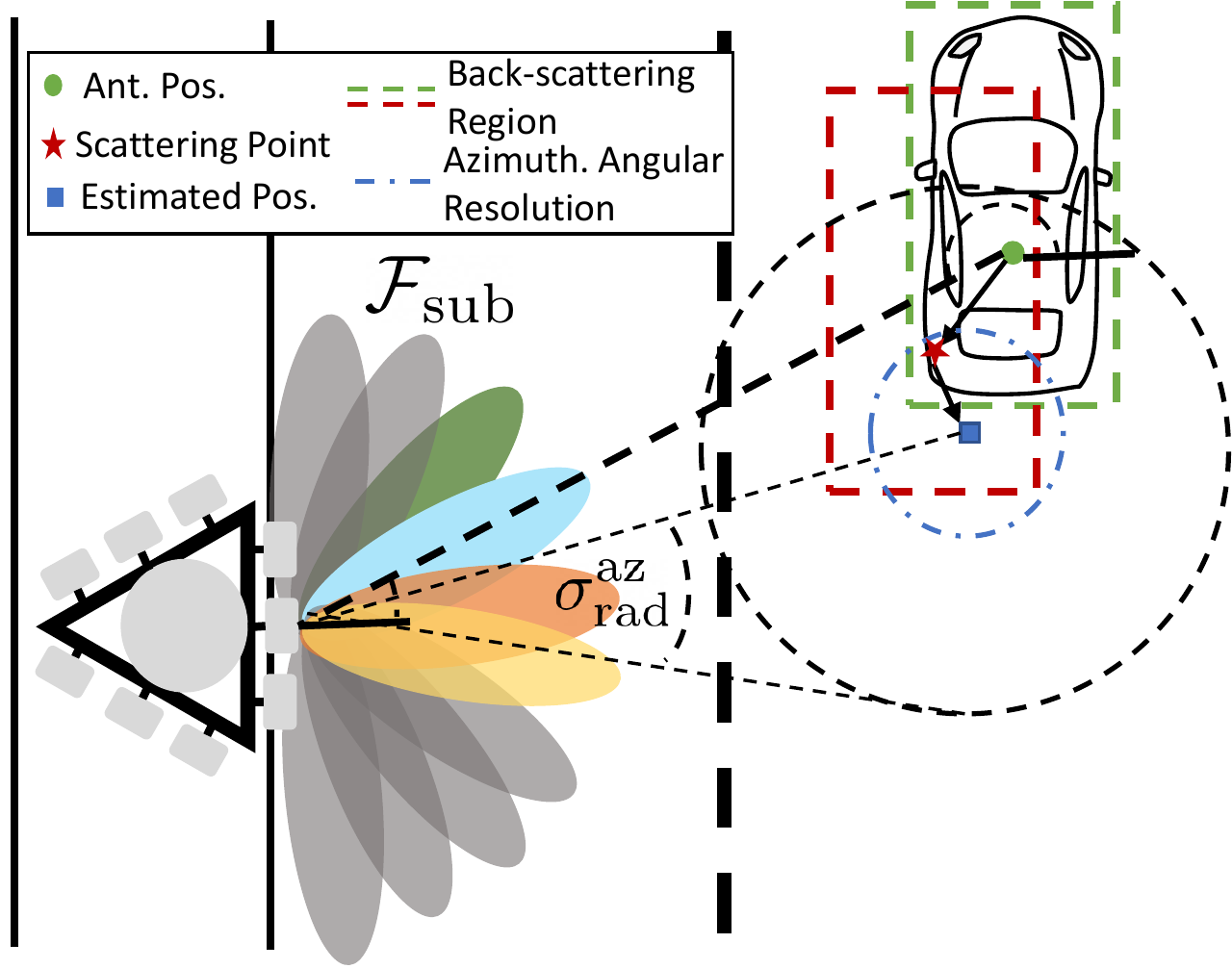}
\caption{Beam training scheme at the BS: Radar-assisted method, the reduced codebook depends on the measurements and scatter point uncertainty $\sigma^{az}_{rad}$.}
\label{fig:rad}
\end{figure}

\subsection{Back-scattering modelling for BM}
According to previous evidences, the radar-estimated antenna position in global Cartesian coordinates can be modelled as 
\begin{equation}\label{eq:radar_estimated_pos_Cart}
\begin{split}
    \hat{\mathbf{p}}_{\mathrm{rad}}  =  \begin{bmatrix}\hat{x}_\mathrm{rad}\\
    \hat{y}_\mathrm{rad}\\
    \hat{z}_\mathrm{rad}\end{bmatrix} = \mathbf{p} + \mathbf{b}(\mathbf{p})+ \mathbf{e}_{\mathrm{res}} + \mathbf{e}_{\mathrm{scatt}},
\end{split}
\end{equation}
where: \textit{(i)} $\mathbf{e}_{\mathrm{res}}$ is the estimation error due to the radar resolution cell, expressed in global Cartesian coordinates; \textit{(ii)} $\mathbf{e}_{\mathrm{scatt}}$ is the uncertainty on the position of the strongest back-scattering point detected by the MIMO radar, i.e., the \textit{size} of the back-scattering region sensed by the radar; \textit{(iii)} $\mathbf{b}(\mathbf{p})$ is a position- and vehicle-dependent bias term shown in Fig.~\ref{fig:radar_measurement} accounting for the shift between the antenna and the scattering centroid. The latter term depends, as discussed before, on the geometry of the radar acquisition and of the size of the target (e.g., car vs truck) and it can be possibly calibrated a-priori.
For the purpose of BM, we include the bias effect into the the size of the back-scattering region, i.e., 
\begin{equation}\label{eq:radar_estimated_pos_Cart_noscatt}
\begin{split}
    \hat{\mathbf{p}}_{\mathrm{rad}} = \mathbf{p} + \mathbf{e}_{\mathrm{res}} + \widetilde{\mathbf{e}}_{\mathrm{scatt}},
\end{split}
\end{equation}
%
where we model the mismatch between the back-scattering point on VE and the true antenna position as as a uniform random variable distributed over the volume of the car, i.e., $\widetilde{\mathbf{e}}_{\mathrm{scatt}}\sim\mathcal{U}(\mathbf{0},\widetilde{\mathbf{C}}_{\mathrm{scatt}})$. The covariance matrix is 
\begin{equation}
    \widetilde{\mathbf{C}}_{\mathrm{scatt}}= \frac{1}{12}\begin{bmatrix}L_\mathrm{VE}^2 & 0 & 0\\
    0 & W_\mathrm{VE}^2& 0\\
    0 & 0 & H_\mathrm{VE}^2\end{bmatrix}
\end{equation}
where we assume that a vehicle of length $L_\mathrm{VE}$, width $W_\mathrm{VE}$ and height $H_\mathrm{VE}$ is oriented along $x$, $y$ and $z$ axis, respectively.

The composite covariance matrix of the radar acquisition, necessary to define the BM set $\mathcal{F}_{\mathrm{sub}}$ at the BS in~\eqref{eq:BS_codebook_rad}, is
\begin{equation}\label{eq:Cart_2_sphe}
    \mathbf{C}^u_{\mathrm{rad}} = \mathbf{C}^u_{\mathrm{res}}+\widetilde{\mathbf{C}}^u_{\mathrm{scatt}}
\end{equation}
where $\widetilde{\mathbf{C}}^u_{\mathrm{scatt}} = \mathbf{T}\mathbf{R} \,\widetilde{\mathbf{C}}_{\mathrm{scatt}}\mathbf{R}^\mathrm{T}\mathbf{T}^\mathrm{T}$ is the inverse covariance mapping from global to local spherical coordinates. Figure~\ref{fig:rad} illustrates the procedure.
Finally, the angular codeooks at the BS for the radar acquisition are obtained from 
$[\mathbf{C}^u_{\mathrm{rad}}]_{(2,2)} = \left( \sigma^\mathrm{az}_{\mathrm{rad}}\right)^2$ and $[\mathbf{C}^u_{\mathrm{rad}}]_{(3,3)} = \left( \sigma^\mathrm{el}_{\mathrm{rad}}\right)^2$ as in~\eqref{eq:BS_codebook_rad}. The same then applies to the MT BM method.

\section{Numerical Results and Discussion}\label{sect:results}

\begin{figure*}[!t]
    \centering
     \subfloat[][]{\includegraphics[width=0.65\columnwidth]{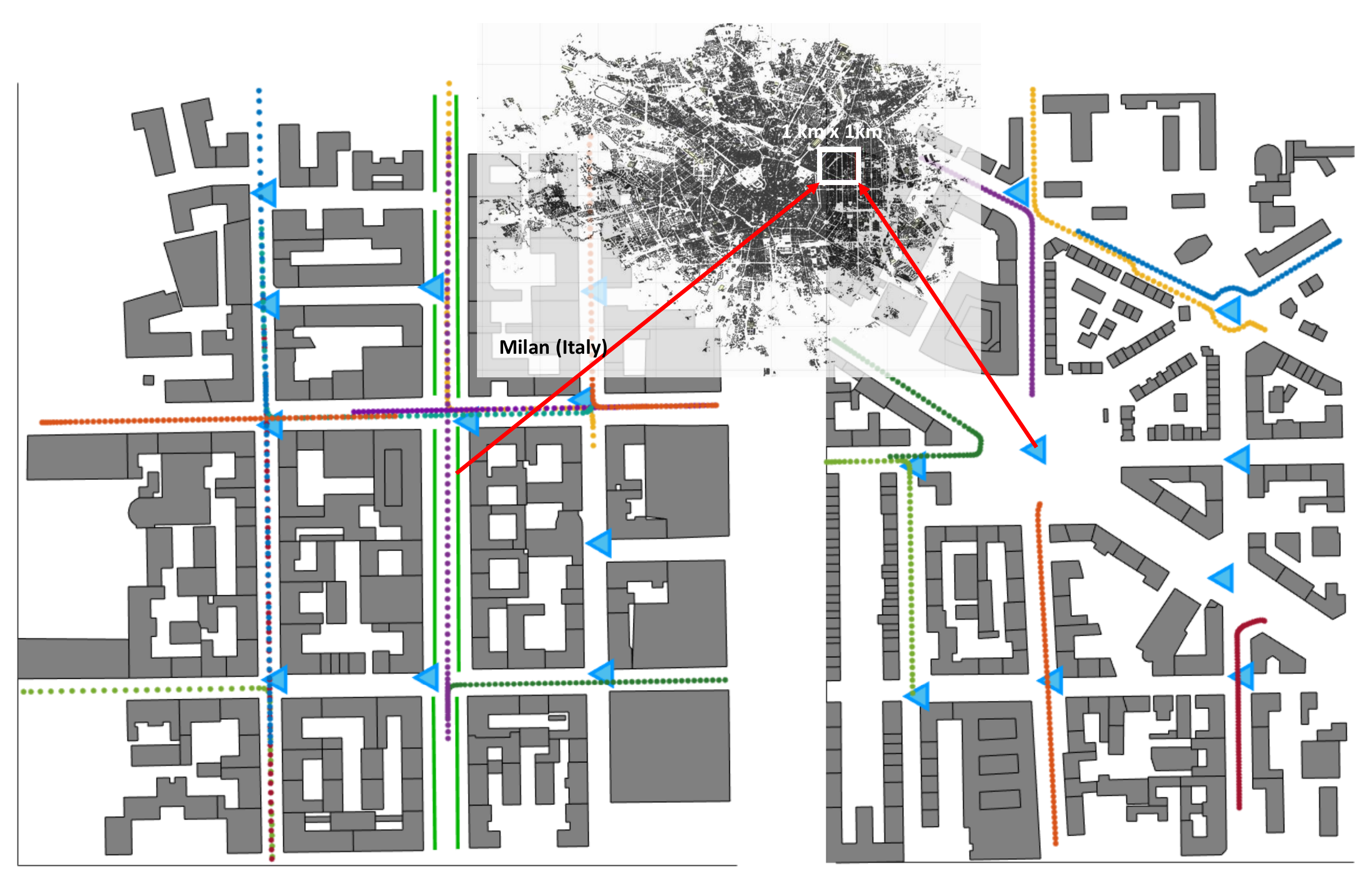}\label{fig:Milano}}
    \subfloat[][]{\includegraphics[width=0.3\columnwidth]{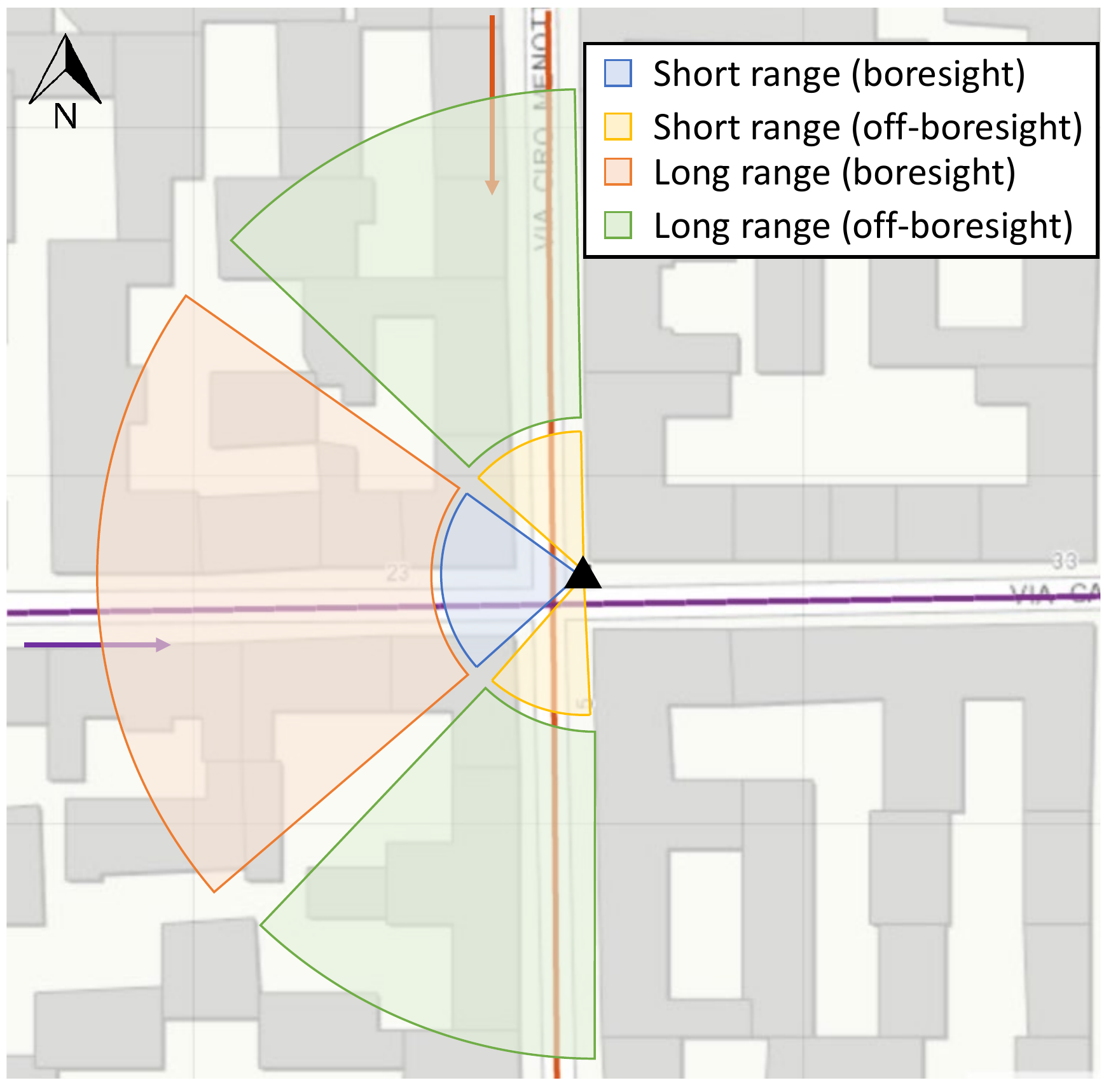}\label{fig:scenario_zone}}
    \caption{ a) simulated trajectories over regions from urban city of Milan and b) definition of the communication regions.}
\end{figure*}
\begin{table}[t!]
    \centering
    \caption{Simulation Parameters}
    \begin{tabular}{l|c}
    \toprule
    \textbf{Parameter} & \textbf{Value(s)}\\
    \hline
        Comm. carrier freq. &  $28$ GHz\\
        N. of BS antenna $N_\mathrm{BS}$ & 16 $\times$  6 \\
        N. of VE antenna $N_\mathrm{VE}$ & 8 $\times$  4 \\
        Single pair test time $T$ & $62.5$ $\mu$s \\
        Tx power $\sigma_s^2$ & $-10$ dBm\\
        Rx noise power $\sigma_n^2$ & $-102$ dBm\\
        SNR threshold $\gamma_\mathrm{thr}$ & 10 dB \\
        VE activity factor $\varepsilon$ & $0.1,\, 0.5,\,1 $ \\
        \hline
        Radar carrier freq. &  $77$ GHz\\ 
        Radar measurement rate $R_\mathrm{rad}$ &  $100$ Hz\\
        Radar Bandwidth &  $1$ GHz\\
        \hline
        Pos. measurement rate $R_\mathrm{pos}$ & $10$ Hz\\
        \bottomrule
        \end{tabular}
        \label{tab:SimParam}
        \end{table}

%
\begin{table}[t!]
    \centering
    \caption{BM Configurations}
    \begin{tabular}{c|c||c|c||c|c}
    \toprule
    \multicolumn{2}{c||}{\textbf{Radar}}&\multicolumn{2}{c||}{\textbf{Position}}&\multicolumn{2}{c}{\textbf{Gradient Search}}\\
    
    \hline
     & \textbf{$N_{\mathrm{rad}}$} &  & $\sigma$ [m]&  & \textbf{$K$}\\
    \hline
	R1 & $32\times4$ & P1 & $1$  & G1 & $1$\\ 
	R2 & $128\times4$  & P2 & $2.5$ & G2 & $2$\\ 
	R3 & $16\times8$ & P3 & $5$ & G3 & $3$\\
	R4 & $64\times8$ & & & G4 & $4$\\ 
	R5 & $8\times16$ & & &\\ 
	R6 & $32\times16$ & & &\\
	\bottomrule
	\end{tabular}
\label{tab:Configuration}
\end{table}

\begin{figure*}[!t]
    \centering
    
    \subfloat[][]{\includegraphics[width=0.5\columnwidth]{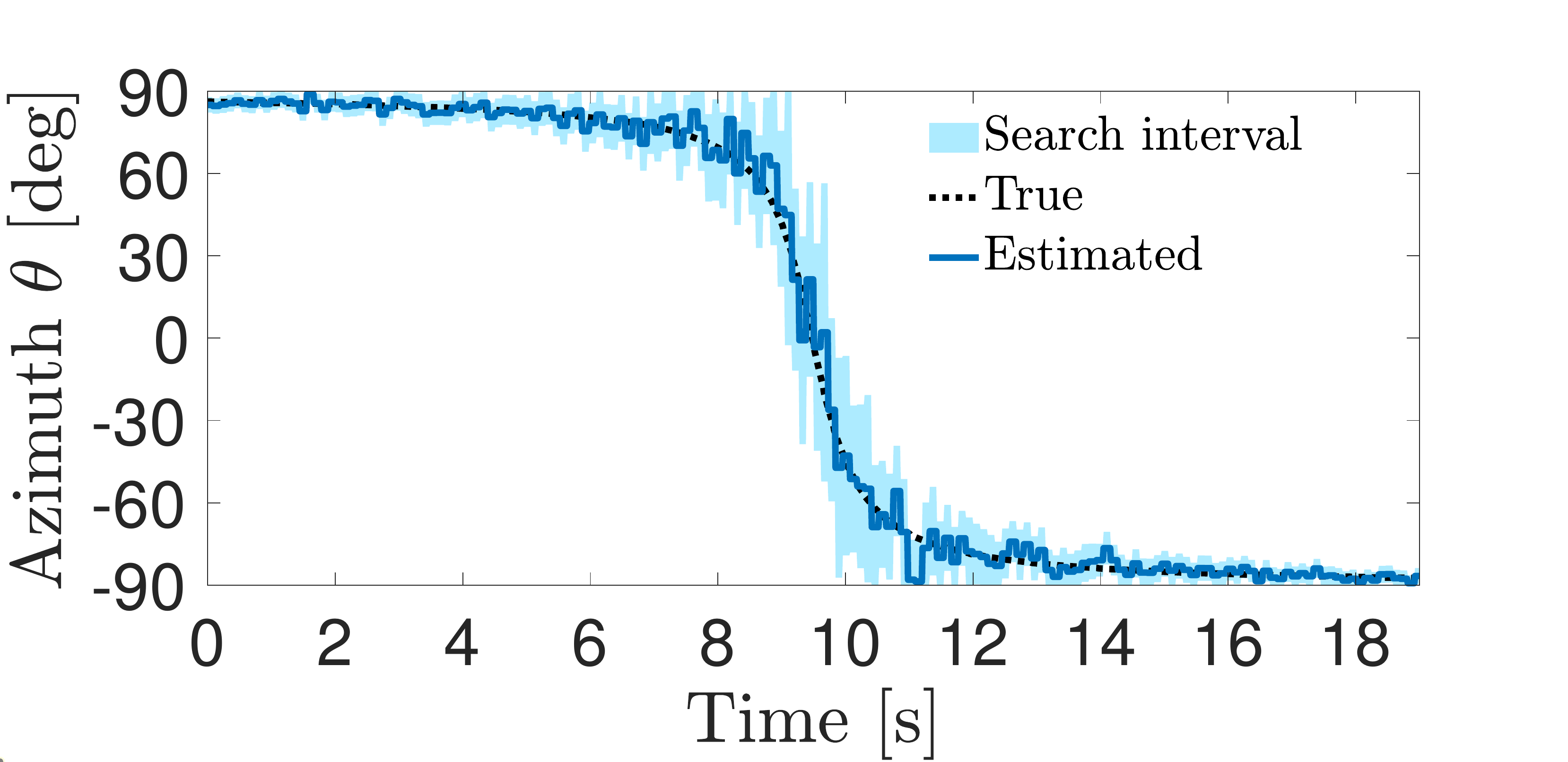}\label{subfig:angles_pos_az}}
    \subfloat[][]{\includegraphics[width=0.5\columnwidth]{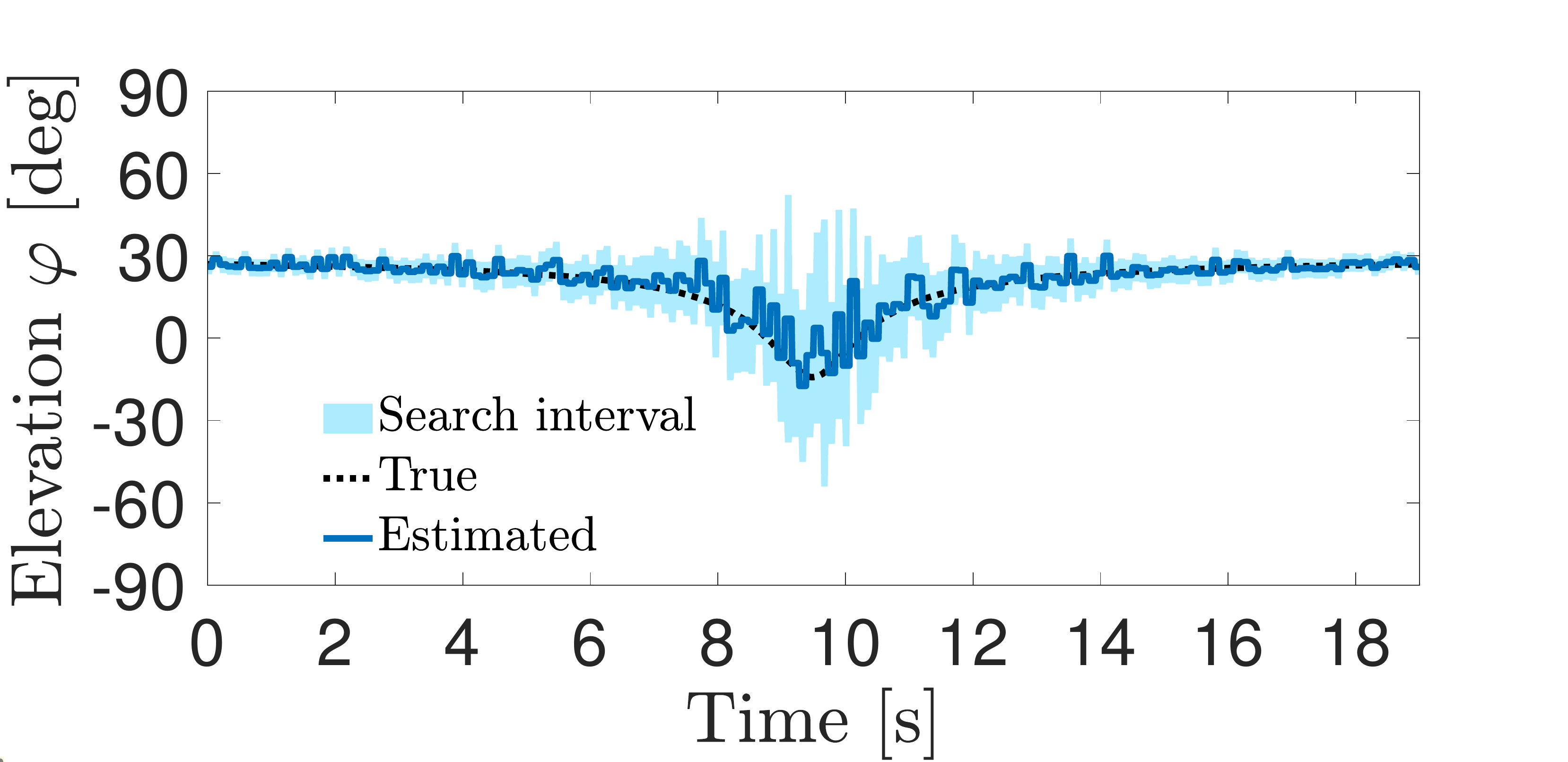}\label{subfig:angles_pos_el}}\\

    \subfloat[][]{\includegraphics[width=0.5\columnwidth]{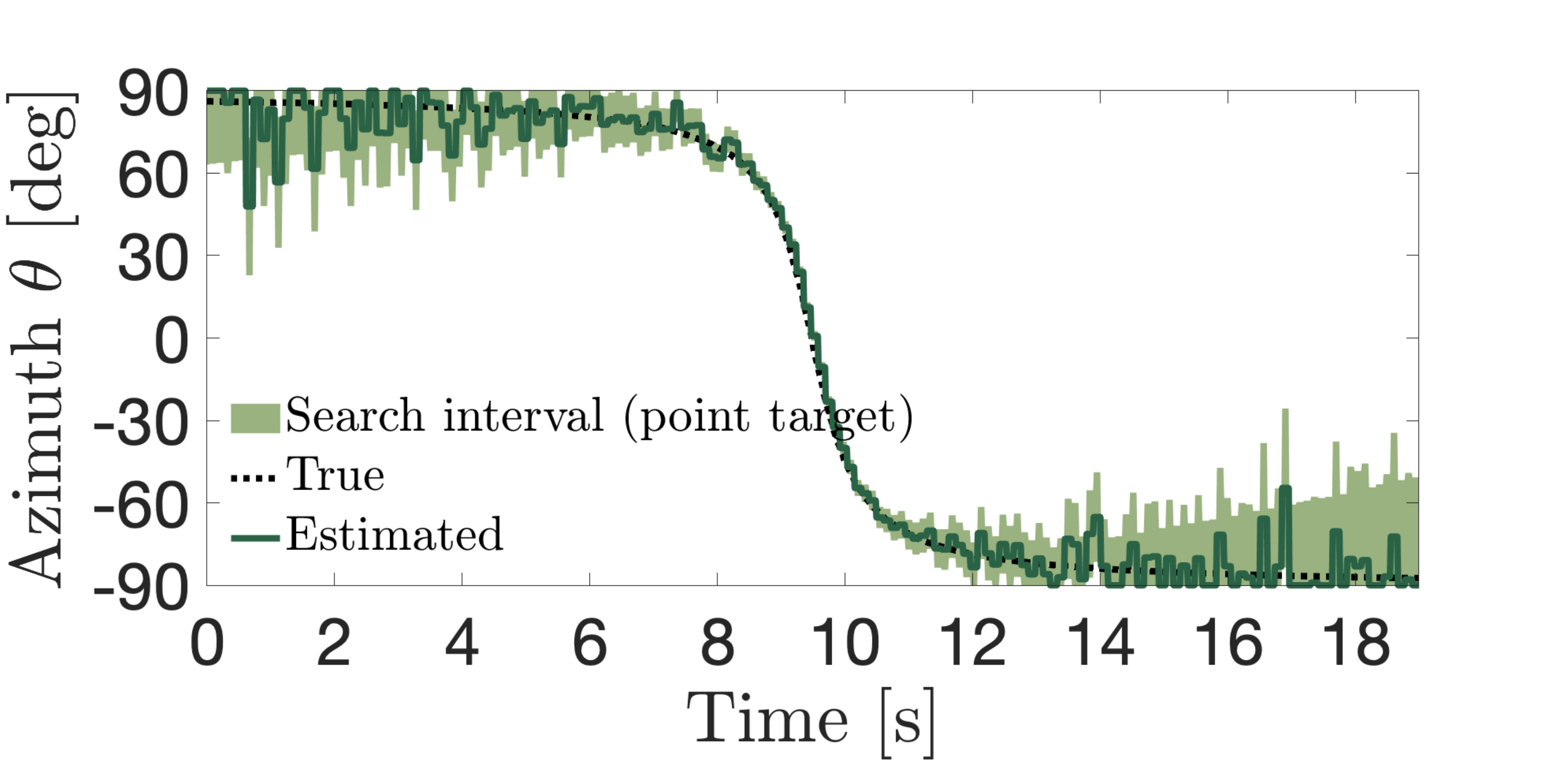}\label{subfig:angles_rad_az_point}}
    \subfloat[][]{\includegraphics[width=0.5\columnwidth]{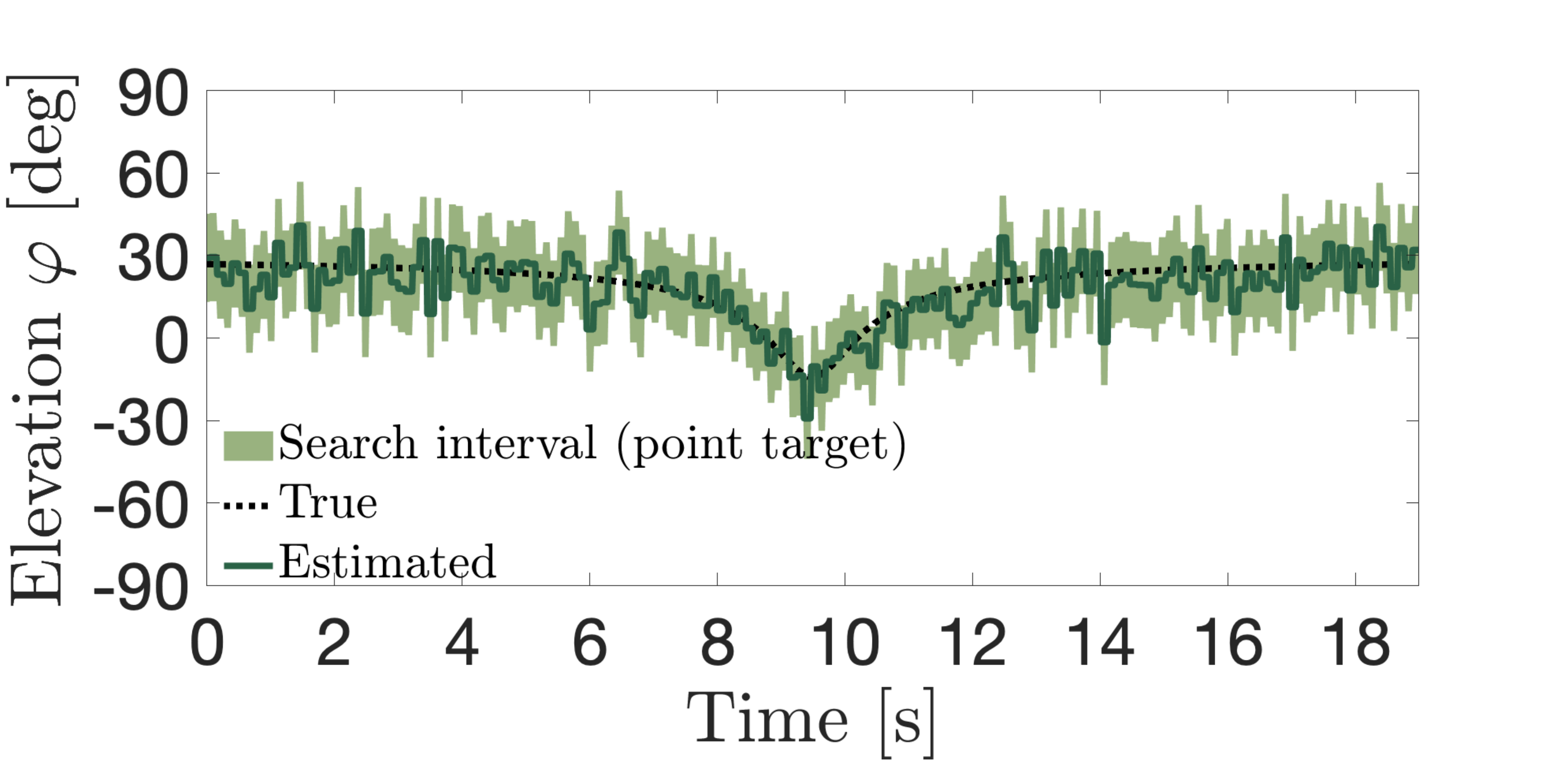}\label{subfig:angles_rad_el_point}}\\
    
        \subfloat[][]{\includegraphics[width=0.5\columnwidth]{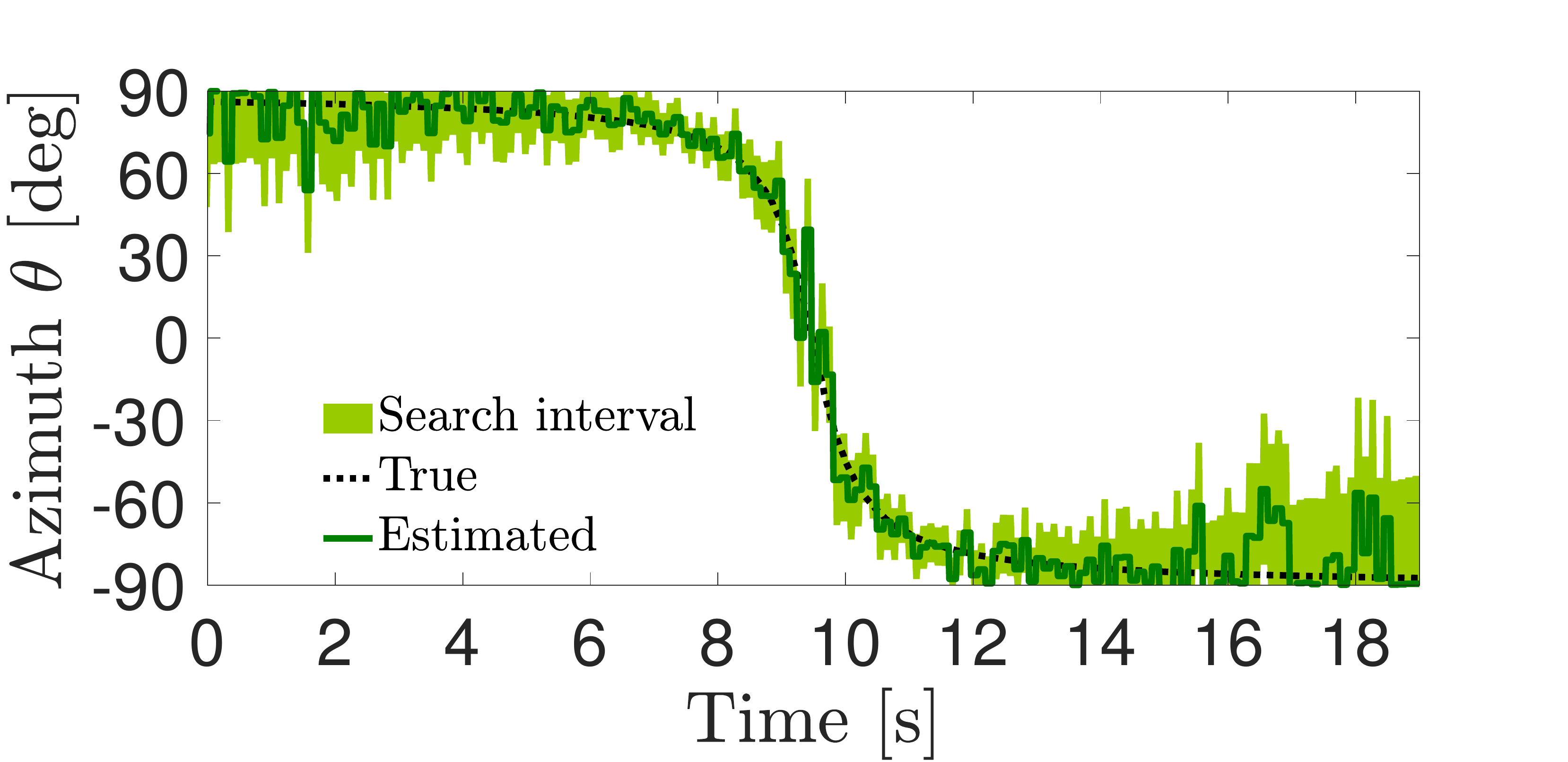}\label{subfig:angles_rad_az}}
    \subfloat[][]{\includegraphics[width=0.5\columnwidth]{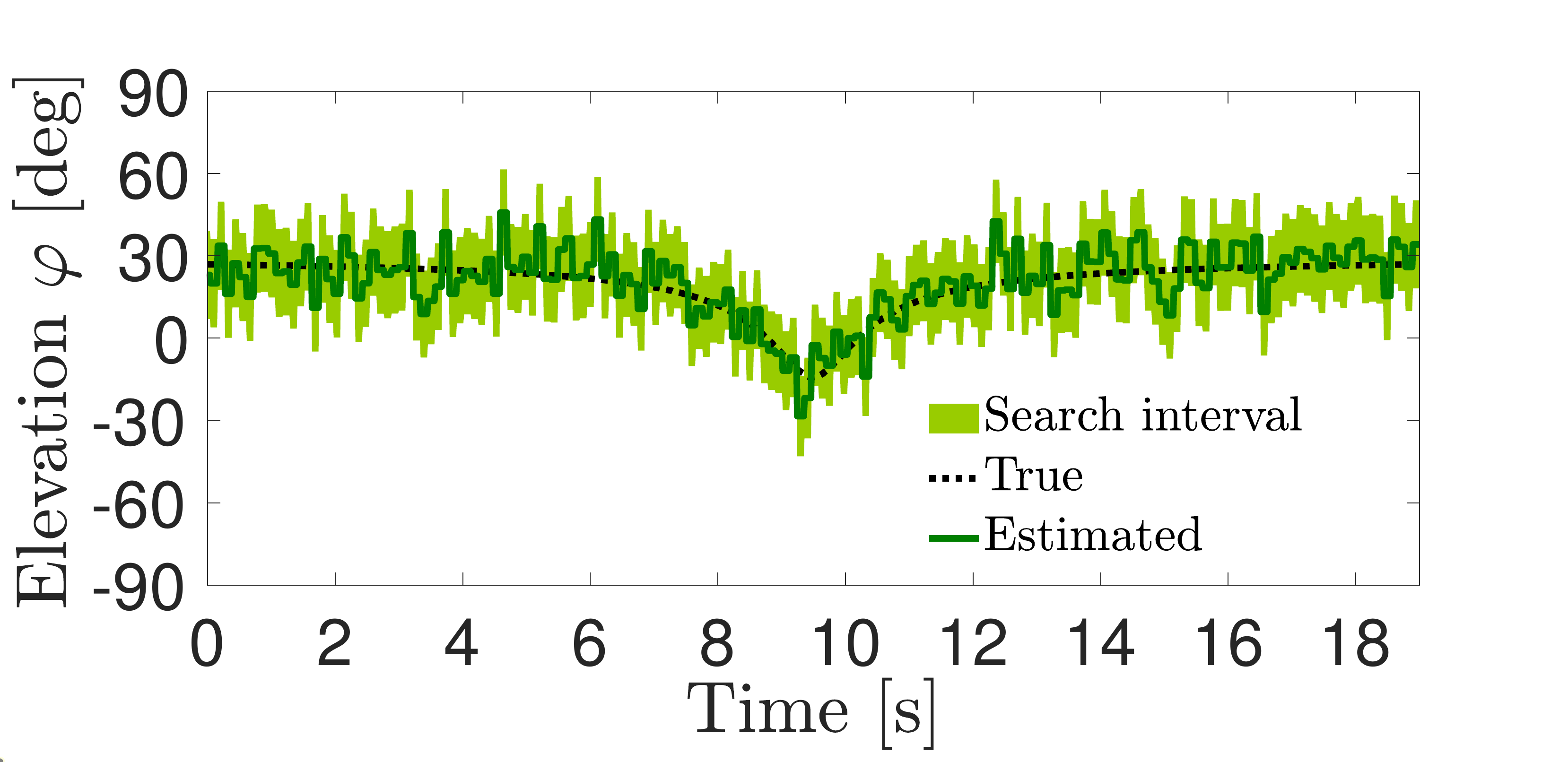}\label{subfig:angles_rad_el}}\\
    
    \caption{Evolution of the true and estimated azimuth and elevation angles in the North-South trajectory in Fig.~\ref{fig:scenario_zone}: (\ref{subfig:angles_pos_az}-\ref{subfig:angles_pos_el}) positioning systems at the VE; (\ref{subfig:angles_rad_az}-\ref{subfig:angles_rad_el}) radar at the BS with offset $\mathbf{b}(\mathbf{p})$, and (\ref{subfig:angles_rad_az_point}-\ref{subfig:angles_rad_el_point}) ideal radar at the BS (with point scatter assumption). The angular search areas, defining the tested codebooks at the BS, is also reported in shaded colors.}
    \label{fig:angles}
\end{figure*}

\begin{figure*}[!t] 
\centering
\subfloat[\label{subfig:TrTime_shortrange_front}]{\includegraphics[width=0.5\columnwidth]{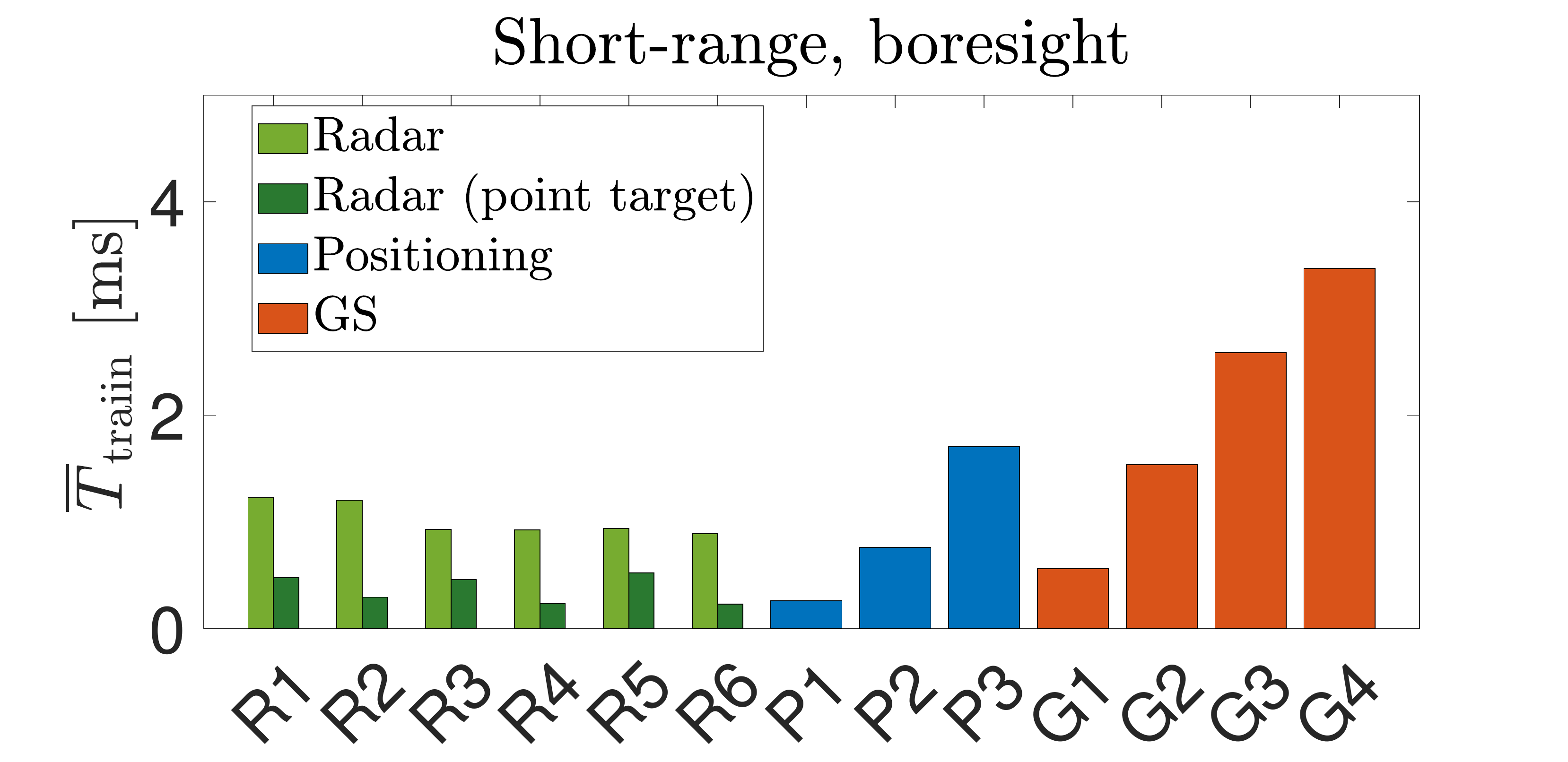}}
\subfloat[\label{subfig:TrTime_shortrange_lateral}]{\includegraphics[width=0.5\columnwidth]{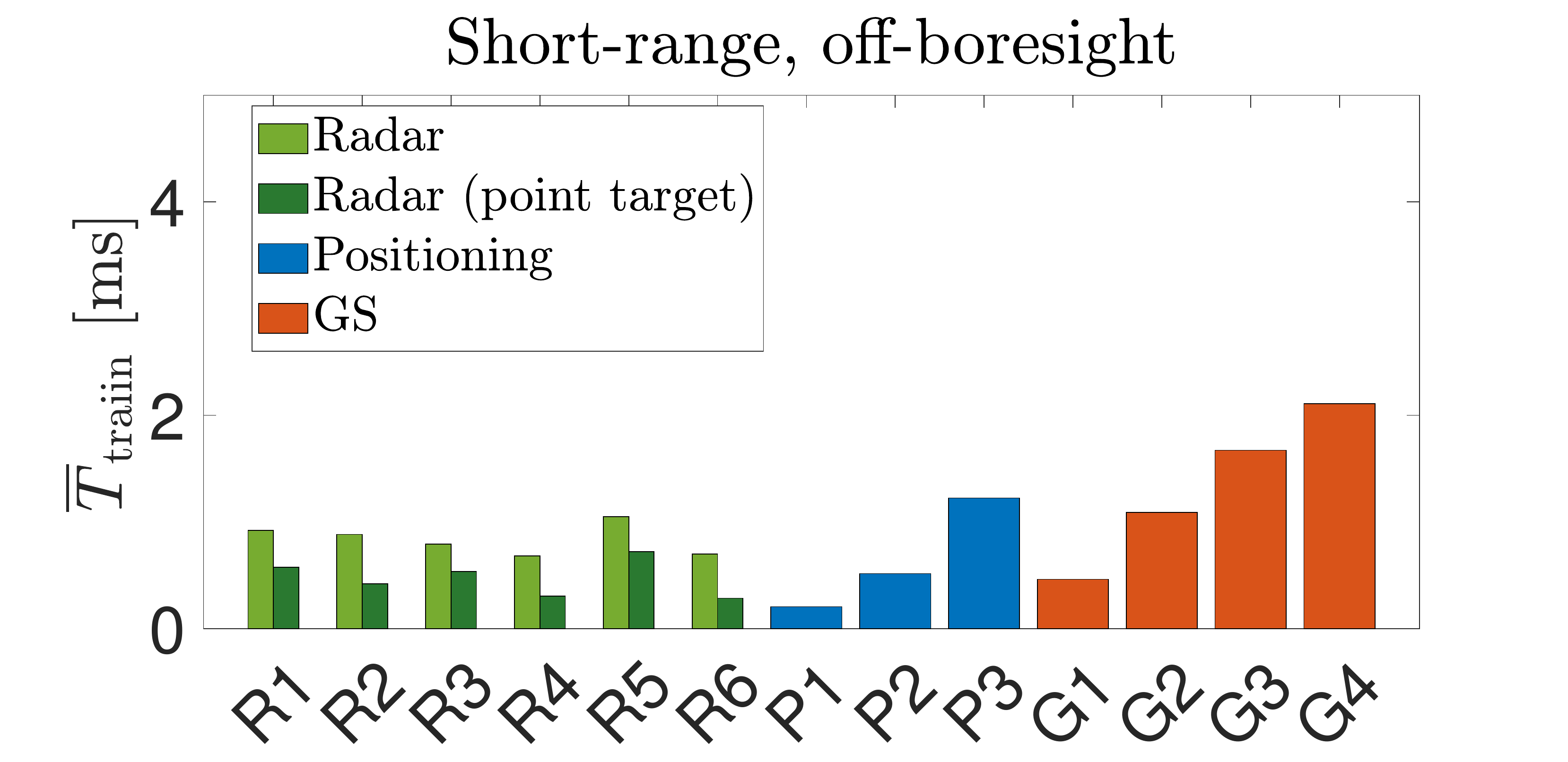}}\\

\subfloat[\label{subfig:TrTime_longrange_front}]{\includegraphics[width=0.5\columnwidth]{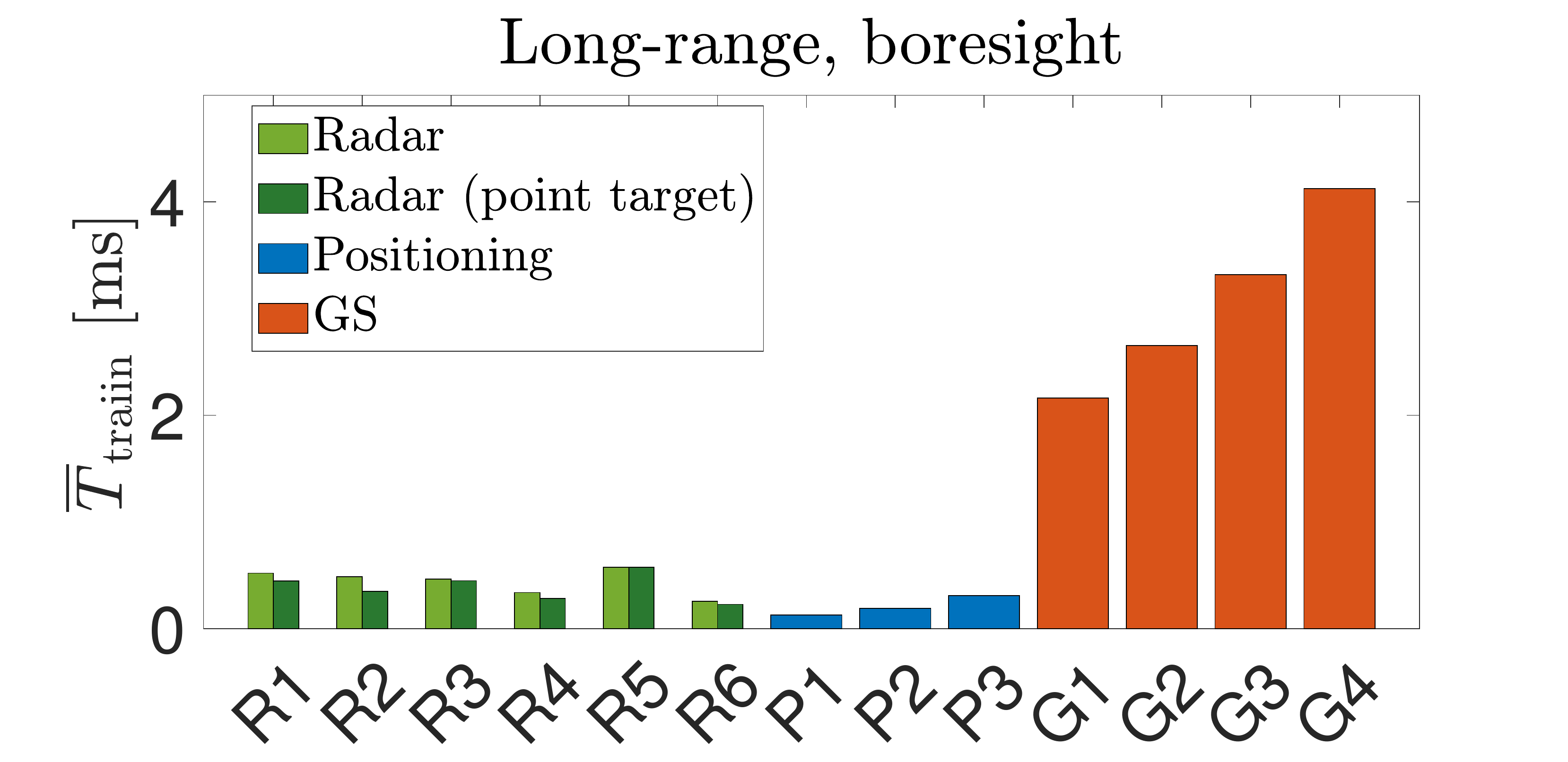}}
\subfloat[\label{subfig:TrTime_longrange_lateral}]{\includegraphics[width=0.5\columnwidth]{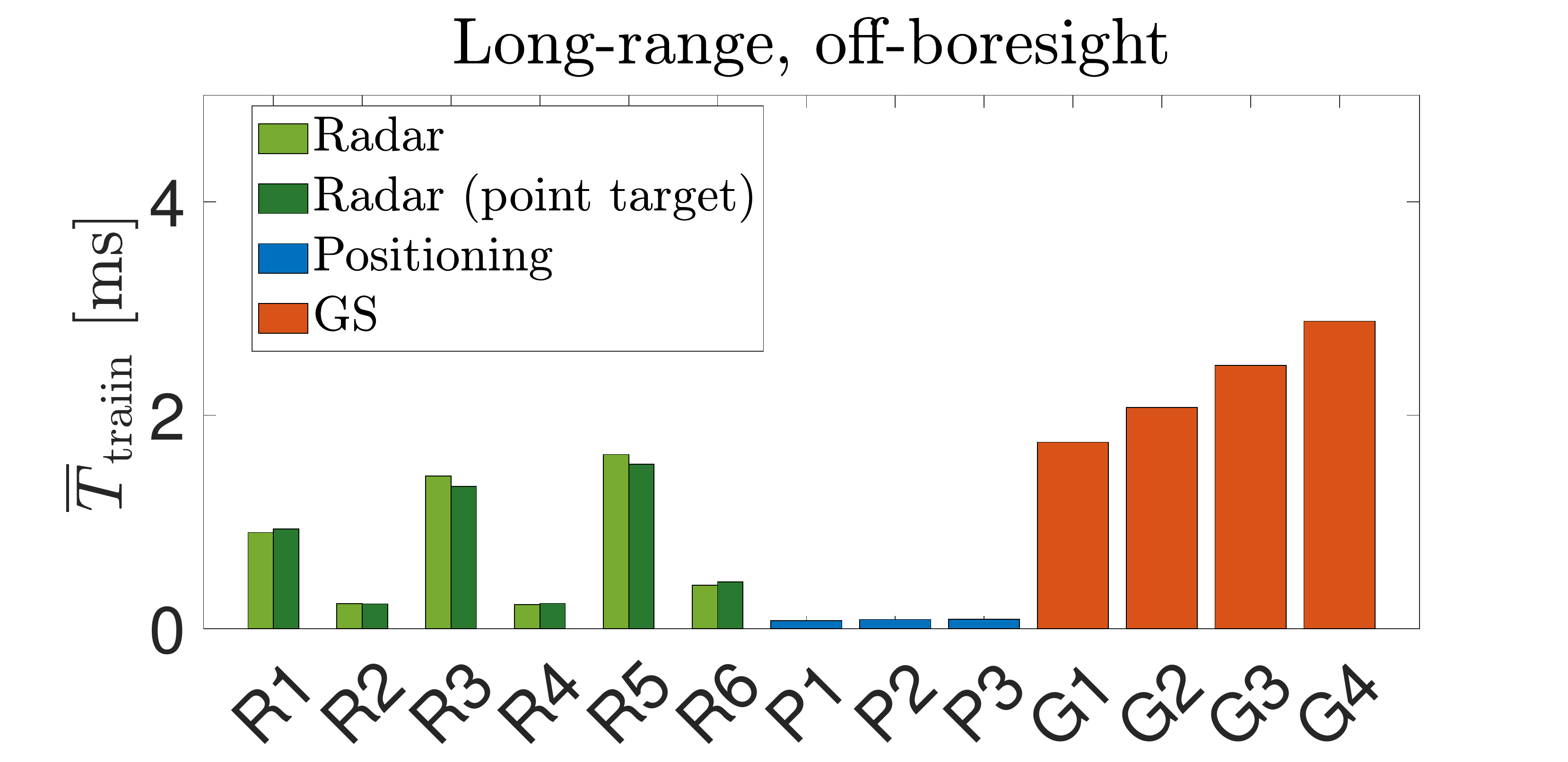}}\\

\subfloat[\label{subfig:TrTime_average}]{\includegraphics[width=0.5\columnwidth]{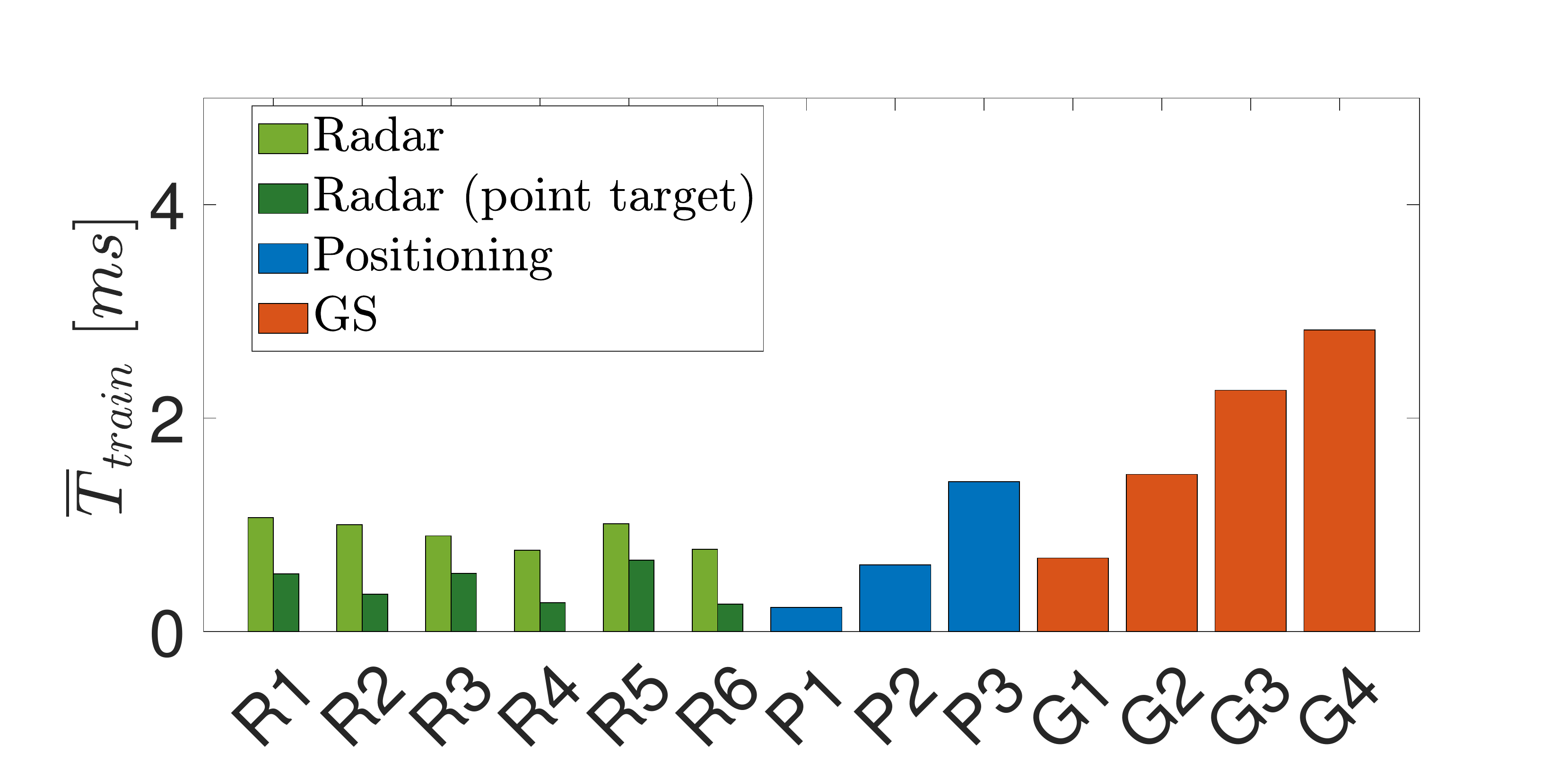}}

\caption{Average training time $\overline{T}_{\mathrm{train}}$ over all the trajectories in Fig.~\ref{fig:Milano} grouped according to the regions of Fig.~\ref{fig:scenario_zone}: (\ref{subfig:TrTime_shortrange_front}) short range, frontal view (boresight), (\ref{subfig:TrTime_shortrange_lateral}) short range, lateral view (off-boresight)
(\ref{subfig:TrTime_longrange_front}) long range, frontal view (boresight) (\ref{subfig:TrTime_longrange_lateral}) long range, and lateral view (off-boresight) of different technologies setting as in Tab~\ref{tab:Configuration}.
}
\label{fig:hist_training_time}
\end{figure*}

\begin{figure*} [!ht]
    \centering
    \subfloat[][]{\includegraphics[width=0.32\columnwidth]{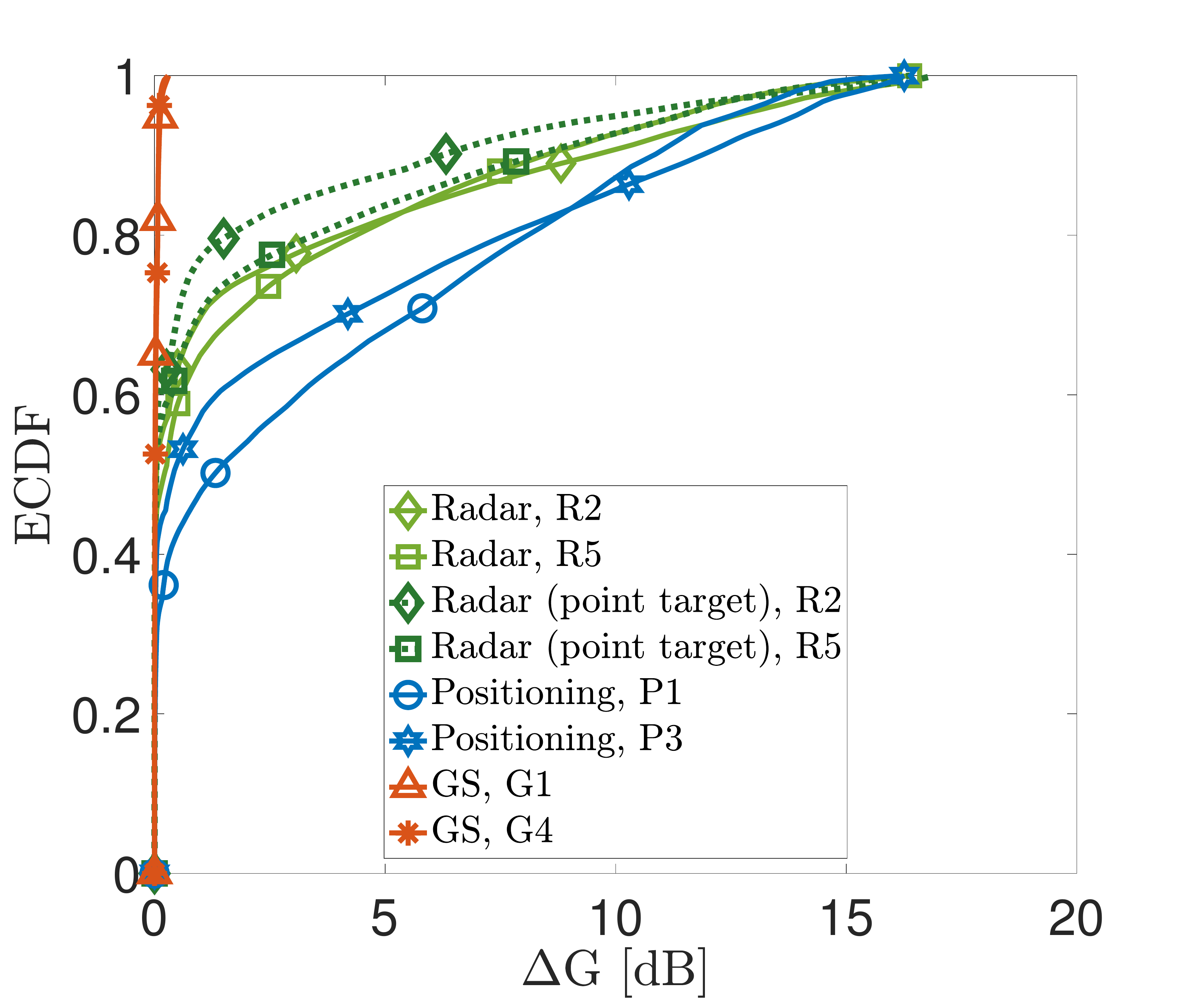}\label{subfig:EDCF_GL_eps1}}
    \subfloat[][]{\includegraphics[width=0.32\columnwidth]{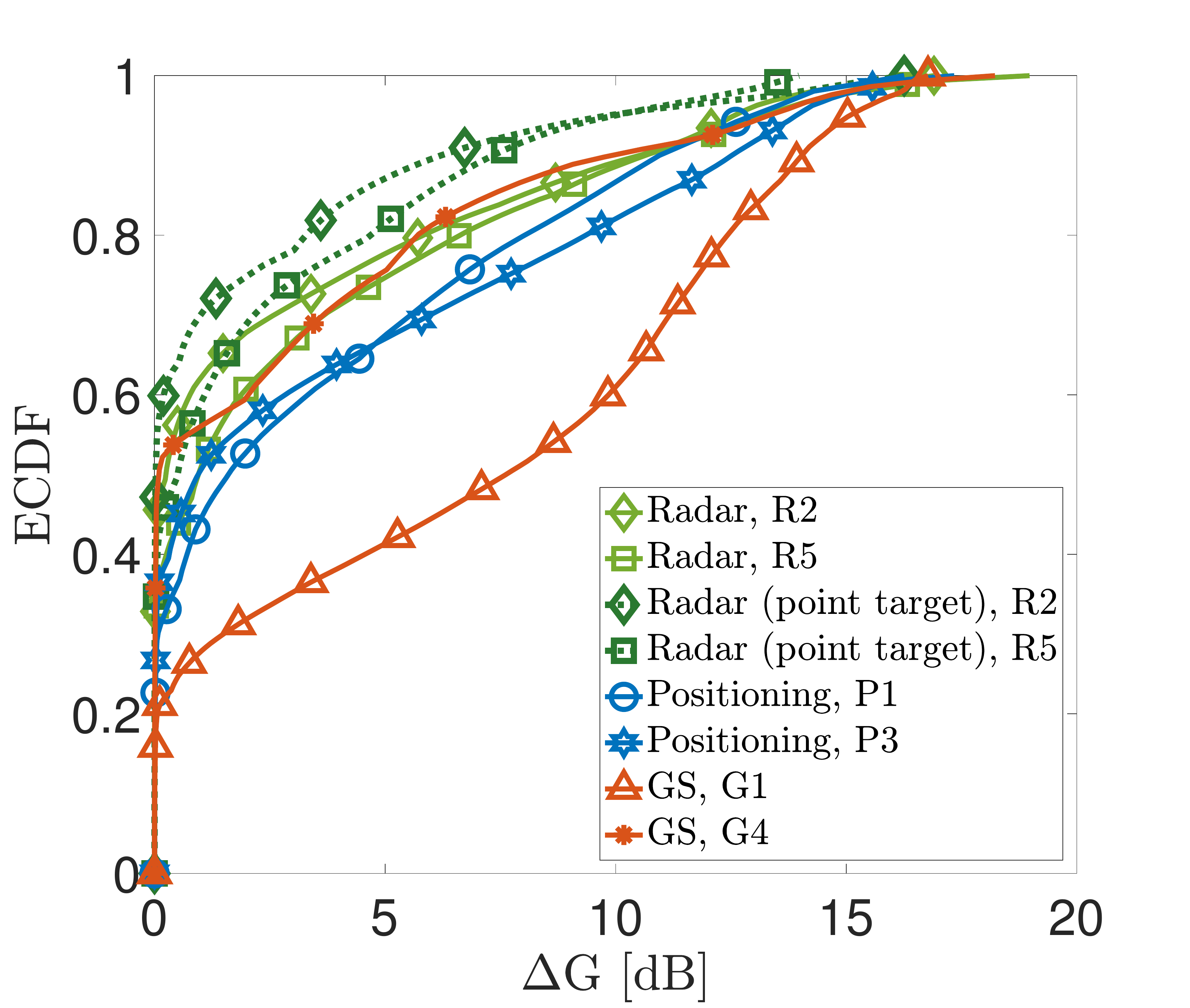}\label{subfig:EDCF_GL_eps05}}
    \subfloat[][]{\includegraphics[width=0.32\columnwidth]{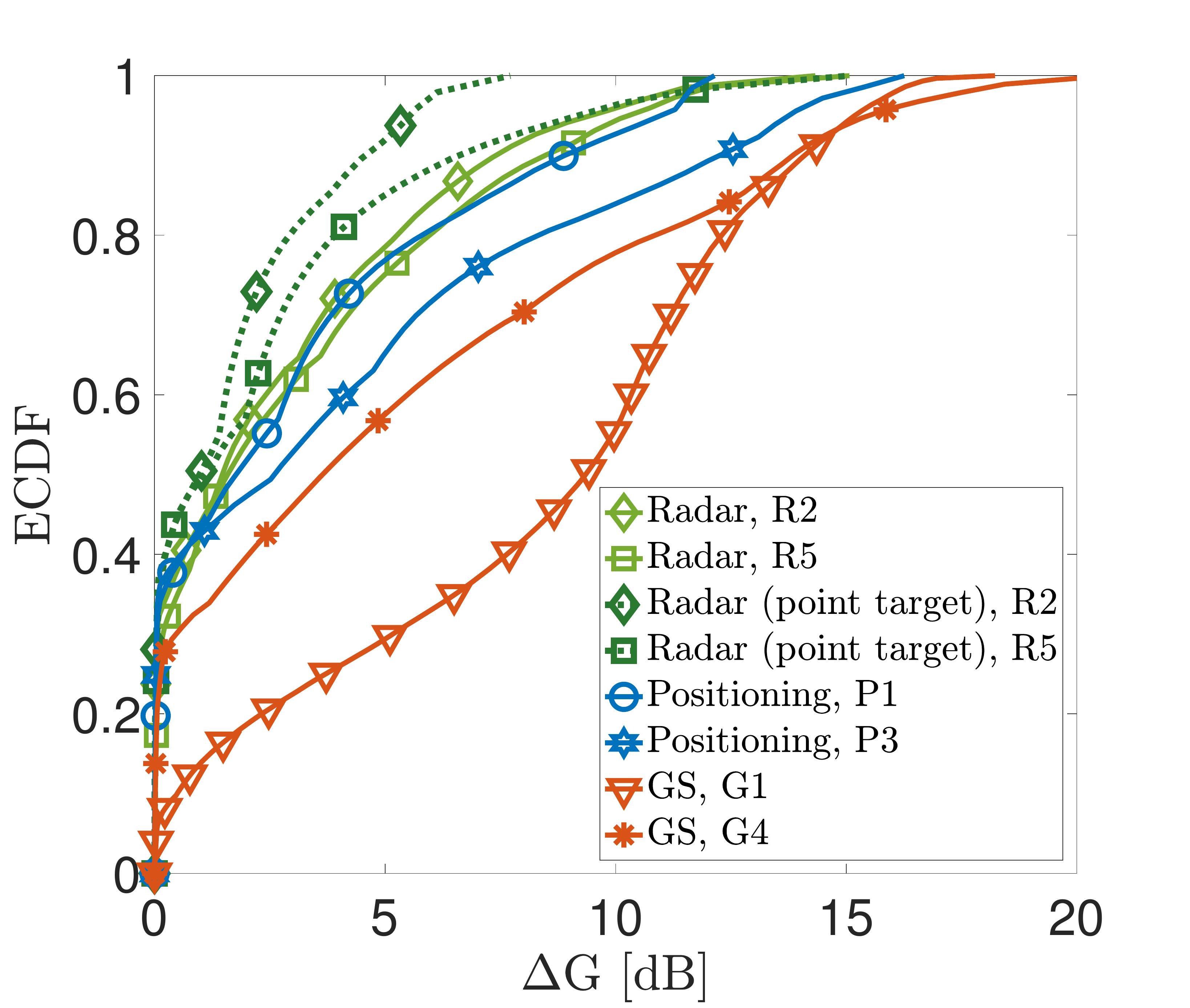}\label{subfig:EDCF_GL_eps01}}
    \caption{ECDF of the beamforming gain loss $\Delta \mathrm{G}$ in case of non-optimal BM ($\hat{\mathbf{f}}\neq\mathbf{f}_{opt}$) varying the activity factor $\varepsilon$: (a) $\varepsilon=1$, (b) $\varepsilon=0.5$, and (c) $\varepsilon=0.1$.}
    \label{fig:ECDF_GL}
\end{figure*}

The BM methods described in the previous section are compared by means of numerical simulation in terms of instantaneous codebook cardinality, average beam training time at each training instant, and beamforming gain loss. 

The selected scenario is (1 km $\times$ 1 km)from the urban area of Milan (Italy) as depicted in Fig.~\ref{fig:Milano}. The geometry of the environment (building shape and position, road route, etc.) is obtained from OpenStreetMap~\cite{osm}, while VE's trajectories are from SUMO~\cite{SUMO2018}, providing multiple realistic vehicular mobility patterns and speeds. The SUMO output is processed by GEMV$^2$~\cite{GEMV2} to obtain the mmW V2I/I2V time-varying channels~\eqref{eq:channel_matrix_freq} (Section~\ref{sect:system_model}) for each time instant and position along the VE trajectory, according to the latest 3GPP guidelines in~\cite{14rel, LinsICC-c21}. Figure~\ref{fig:Milano} are two exemplified BS deployments over 200 m $\times$ 200 m. In Fig.~\ref{fig:Milano}, the BSs' panels are located at each crossroads and at each roundabout, at $7$ m height, tilted by $10$ deg downward. 
Several reference VE trajectories are considered. In most of them, the VEs experience a line of sight I2V downlink channel. 
The trajectories are selected so that to span all the azimuth angles and ranges and to test the full elevation capabilities of the radar.
For the sake of clarity in the results, we classify them according to different communication regions within the coverage area of the BSs (as exemplified in Fig.~\ref{fig:scenario_zone}), to highlight different behaviors of different BM methods. 
System parameters are summarized in Table~\ref{tab:SimParam}. The communication setting is chosen according to the latest 3GPP recommendations~\cite{TS_38213}, while radar parameters are based on the experimental hardware set-up in Sec.~\ref{sect:radar_backscattering}~\cite{TI_ref_MMWCAS}.
The configurations for the different BM methods are reported in Table~\ref{tab:Configuration}, properly labeled to be used in the results. Radar setups are chosen to balance azimuth and elevation resolution, dividing between less expensive setups (R1, R3 and R5) and more expensive ones (R2, R4, R6). Three different positioning systems at the VE are considered, with variable accuracy. Blind GS is simulated accounting of the $K$ value. $K=1$ stands for a system that is supposed to track the angle variation with low overhead, while $K=4$ demands an higher search interval $\mathcal{F}_{\mathrm{sub}}$.
In the BM implementation, whenever the search for the optimum BS beam in $\mathcal{F}_\mathrm{sub}$ does not lead to a satisfactory SNR, i.e., $\gamma(\hat{\mathbf{f}},\mathbf{w}_{opt})< \gamma_\mathrm{thr}$, the BM procedure continues on neighbor beams of $\mathcal{F}_\mathrm{sub}$ until the threshold SNR $\gamma_\mathrm{thr}$ is matched.  
Moreover, we simulate different activity factors $\varepsilon$ for the VE, namely the percentage of time in which the VE is continuously communicating to the BS~\cite{activityfactor}. 
If $N_{tot}$ is the total number of data packets  to be transmitted~\cite{3GPP_5Greq2020}, the probability of having $N_{act}$ consecutive transmissions is 
\begin{align}
   \mathbb{P}(N_{act})= \frac{\lambda^{N_{act}}}{N_{act} !} e^{-\lambda_a}
\end{align}
with $\lambda_a=\varepsilon N_{tot}$.
The value of $\varepsilon$ is mainly influenced by the BS resource (time, frequency, and space) allocation scheduling. As shown in the following, the performance of blind BM methods are affected by the activity factor.

\subsection{Codebook cardinality}

Figure~\ref{fig:angles} shows the evolution of true and estimated azimuth and elevation angles of the VE's antenna from the BS's prospective, and the search interval set $\mathcal{F}_{\mathrm{sub}}$ over time, for the North-South reference trajectory in Fig.~\ref{fig:scenario_zone}, considering position-assisted and radar-assisted BM methods. In particular, we compare configuration P3 for the positioning system (accuracy $\sigma=5$ m) and configuration R6 for radar ($\approx$ $3.5$ and $7$ deg azimuth and elevation resolution, with and without considering the point scattering assumption for the VE) to show the impact of the scattering uncertainty. We also show the angular search area, i.e., the instantaneous codebook cardinality. Positioning sensors (Figs.~\ref{subfig:angles_pos_az} and~\ref{subfig:angles_pos_el}) provide an estimation of the VE's antenna location whose accuracy is approximately invariant within the coverage area of the BS. 
Therefore, the number of beams to be tested $|\mathcal{F}_{\mathrm{sub}}|$ when the VE is close to the BS increases. The motivation of such a behaviour is purely geometrical.
Differently, the radar resolution in both azimuth and elevation is maximum in the boresight direction, thus for the VE located in front of the BS, and it degrades for off-boresight angles. If the radar back-scattering point over the VE is assumed to correspond to the antenna location (Figs.~\ref{subfig:angles_rad_az_point} and~\ref{subfig:angles_rad_el_point}), the radar achieves near-ideal performance nearby the BS, being limited to the sole resolution. This effect can be observed in the steep azimuth transition in Figs.~\ref{subfig:angles_rad_az}~and~\ref{subfig:angles_rad_az_point}, corresponding to the time interval in which the VE is travelling in front of the BS. In this case, the codebook cardinality, and the related beam training time \eqref{eq:TrTime}, is minimum. However, as the VE behaves as an extended target (Figs.~\ref{subfig:angles_rad_az} and~\ref{subfig:angles_rad_el}), the radar-assisted BM should naturally enlarge the angular search area to find the optimal beam as for~\eqref{eq:selected_beamformers}. 

\subsection{Beam training time}
The second set of results concerns the average beam training time $\overline{T}_\mathrm{train}$ (see Section~\ref{sect:beam_training_schemes}) at each beam training instant, for all the configurations of radar-assisted, position-assisted and GS BM methods in Table~\ref{tab:Configuration}. The results are averaged over multiple realizations of the two reference trajectories in Fig.~\ref{fig:scenario_zone}, i.e., North-South and East-West, obtained by SUMO varying VE's speed, position on the lane according to traffic constraints. The activity factor of the VE is here set to $\varepsilon=1$, i.e., continuous communication. Fig.~\ref{fig:hist_training_time} shows $\overline{T}_\mathrm{train}$ in the four different regions of the cell depicted in Fig.~\ref{fig:scenario_zone}, i.e., short/long-range and boresight/off-boresight azimuth angles. 
The first observation is that radar- and position-assisted BM methods uniformely outperform GS ones, in each region within the cell, except for configuration G1 in short-range. 

For short-range communications (Figs.~\ref{subfig:TrTime_shortrange_front} and~\ref{subfig:TrTime_shortrange_lateral}, $\rho\leq 30$ m), it is clear that radar- and position-assisted BM methods consistently outperform GS ones except for G1 configuration. The latter is a fast neighbour search around the optimal pointing derived in the previous training instant, that provides remarkable performance only when the VE is seamlessly connected to the BS ($\varepsilon=1$), as shown in the following. All the radar configurations (R1-R6) show an average training time $\leq 1$ ms, comparable to a medium-to-high accuracy positioning system (P2, $\sigma\leq 2.5$ m). When the positioning accuracy in is the order of $1$ m (P1), the only competitive radar configurations are R2, R4 and R6, that provide a good trade-off between azimuth and elevation resolution. Notice that R4-configured radar capable of detecting the true position of the VE's antenna would halve the training time, down to $\approx 0.5$ ms (ideal situation).

In long-range scenarios (Figs.~\ref{subfig:TrTime_longrange_front} and~\ref{subfig:TrTime_longrange_lateral}), GS are always outperformed by positioning systems and radar, regardless of the configuration, due to the penalty introduced by aligning errors (especially for G1) forcing a codebook enlargement, as explained at the beginning of this section. Along the boresight direction, the main difference with Fig.~\ref{subfig:TrTime_shortrange_front} is that the effect of the extended target nature of the VE plays a minor role, and low-accuracy positioning sensors (P3) provide remarkable performance, comparable with the best radar ones ($\overline{T}_\mathrm{train}$ reduces from $\approx 2$ ms to less than $0.5$ ms for P3). For off-boresight angles (Fig.~\ref{subfig:TrTime_longrange_lateral}, $45\leq\vartheta\leq 90$ deg), instead, the advantage of any positioning solution at the VE is clear, even against comparably expensive radar setups such as R2 and R4.   

The average training time $\overline{T}_\mathrm{train}$ over the whole coverage area is reported in Fig.~\ref{subfig:TrTime_average}. Overall, both radar and positioning systems allow to outperform the best GS solution, except for pessimistic positioning performance (P3, $\sigma=5 $ m). All radar configurations achieve comparable average performance, although expensive ones (R2, R4 and R6) allow uniform training times within the cell. In general, radar shall have the same azimuth and elevation resolution to maximize the performance. However, the next generation of positioning systems with sub-meter accuracy could represent a valid alternative whenever the radar is not able to correctly detect the target (as in NLOS situations).

\subsection{Beamforming gain loss}

The beamforming gain loss $\Delta \mathrm{G}$~\eqref{eq:gain_loss} of the different technologies is investigated here. We report the Empirical Cumulative Distribution Function (ECDF) of $\Delta \mathrm{G}$ conditioned to a misalignment event, i.e., $\hat{\mathbf{f}} \neq \mathbf{f}_{opt}$, for three values of the activity factor, $\varepsilon = 1, 0.5, 0.1$. In all cases and for all methods, the probability $\mathrm{P}$ of finding the global optimum is larger than $90\%$. We select, for each BM method, two configuration with complementary performance. For radar, we choose R2 (maximum azimuth resolution) and R5 (maximum elevation resolution), while P1 and P3 for the positioning system and G1 and G4 for GS. 

For $\varepsilon=1$ (Fig.~\ref{subfig:EDCF_GL_eps1}), GS provides optimal performance, with a negligible beamforming penalty, as the tracking of the optimal beam is nearly perfect.
However, for $\varepsilon<1$ (Figs.~\ref{subfig:EDCF_GL_eps05} and~\ref{subfig:EDCF_GL_eps01}), the random interruptions of BS-VE communication make the GS losing the optimal beam, so that the beam search around an obsolete optimal beam leads to a substantial received power reduction. This is more evident for G1 ($K=1$), where fast beam training is compensated of the vulnerability to the lack of connectivity, propagating the error in time. It is worth noting that both radar- and position-assisted BM losses are not evidently affected by the value of $\varepsilon$, since periodic measurements at rates $R_\mathrm{rad}$ and $R_\mathrm{pos}$ provide fresh information, independently of the communication status. Thus, sensor-asssited solution result in a 3 dB more beamforming gain in average than blind solutions. 
The uncertainty on the scattering point on VE (solid green lines) introduces a $2$ dB extra loss in average with respect to the ideal single scattering point assumption (dashed green lines).

\subsection{MT performance}

\begin{figure} [!t]
    \centering
    \subfloat[][]{\includegraphics[width=0.9\columnwidth]{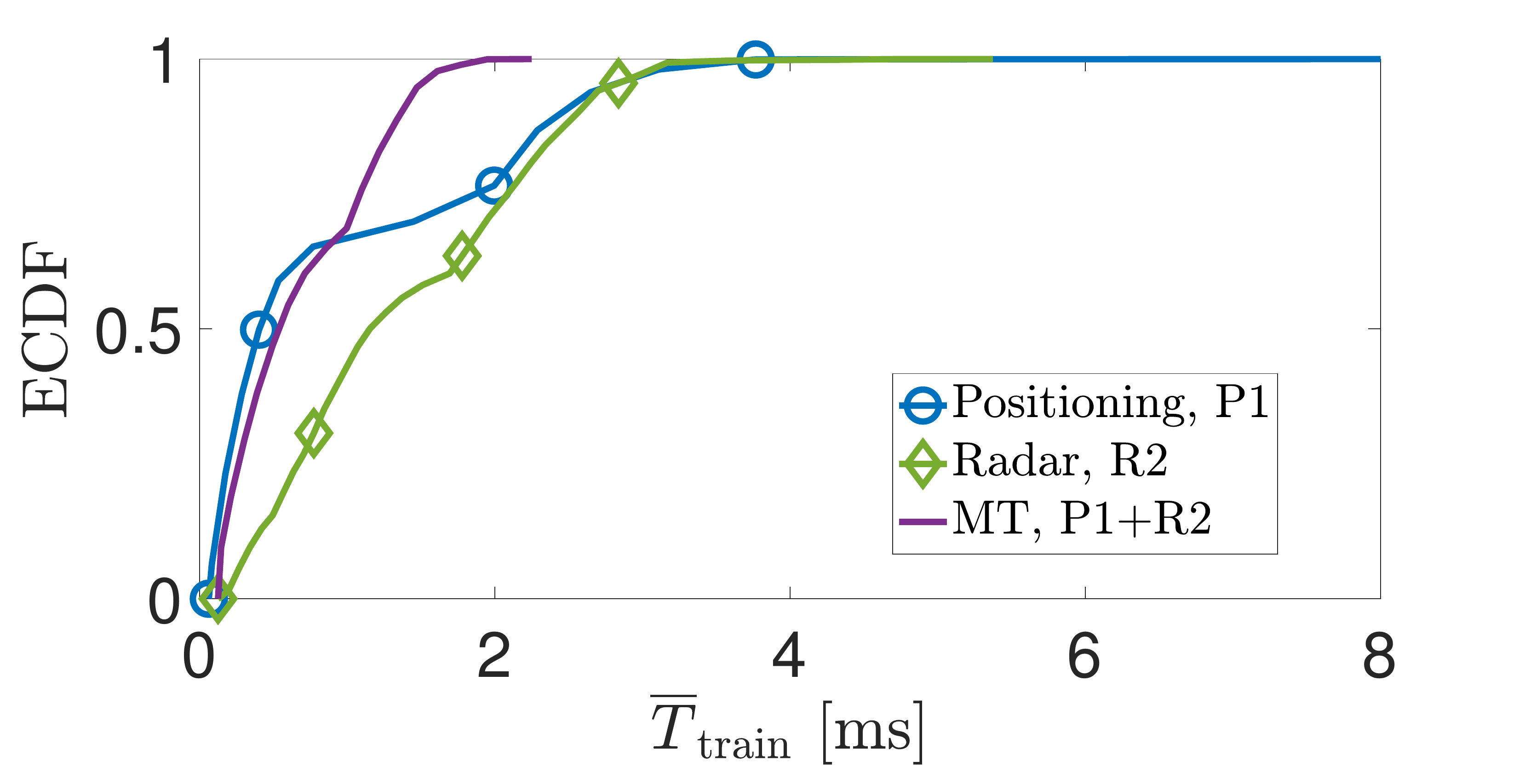}\label{subfig:training_time_MT_ECDF}}\\
    \subfloat[][]{\includegraphics[width=0.9\columnwidth]{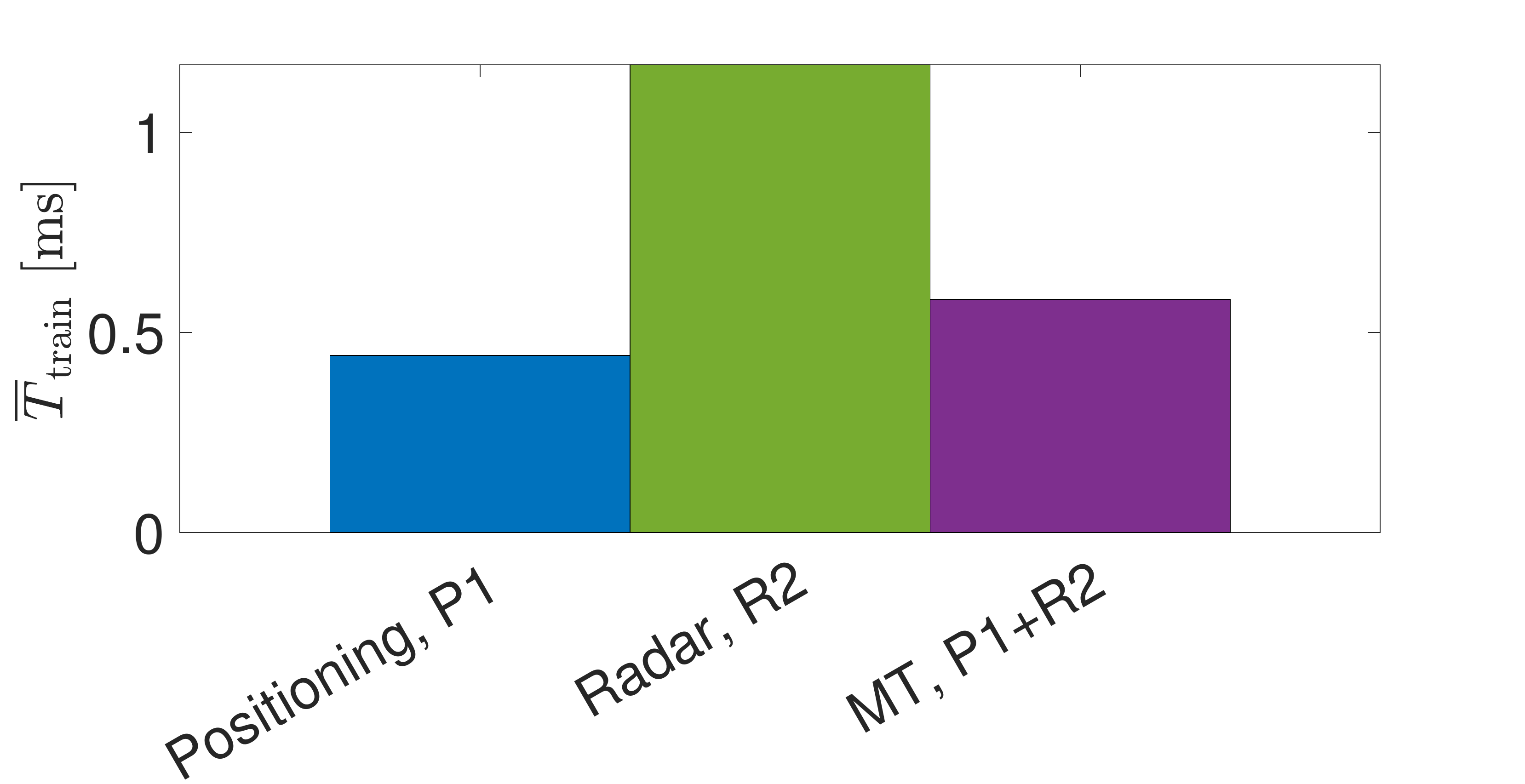}\label{subfig:training_time_MT_barplot}}\\
    \caption{ (a) ECDF and (b) mean value of the beam training time $\overline{T}_{\mathrm{train}}$ for MT BM}
    \label{fig:training_time_MT}
\end{figure}
\begin{figure} [!ht]
    \centering
    \subfloat[][]{\includegraphics[width=0.9\columnwidth]{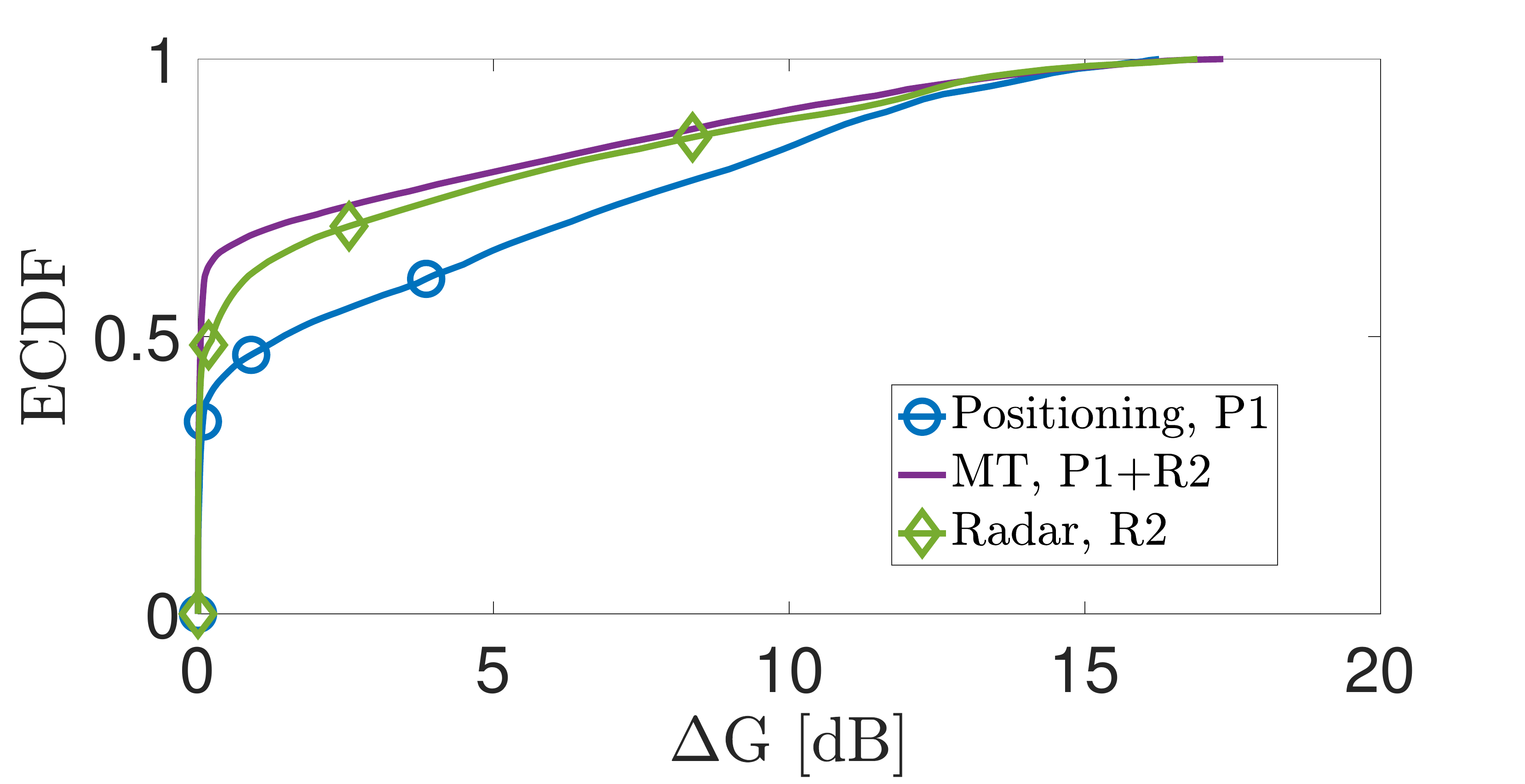}\label{subfig:EDCF_GL_MT_eps01}}\\
    \subfloat[][]{\includegraphics[width=0.9\columnwidth]{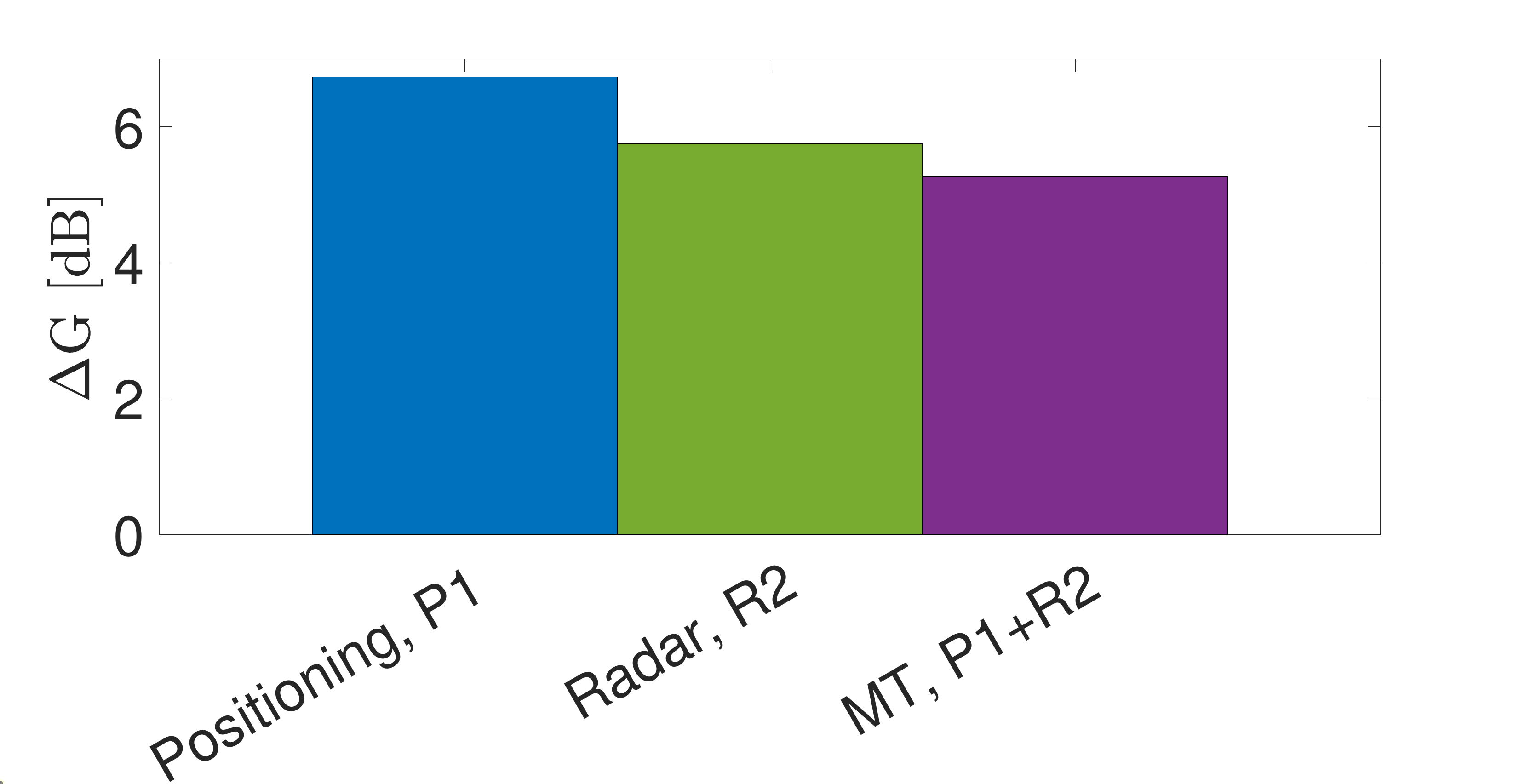}\label{subfig:MTmeanValueDeltaG}}
    \caption{(a) ECDF and (b) mean value of the gain loss $\Delta \mathrm{G}$ with MT BM, considering a VE activity factor $\varepsilon=0.5$}
    \label{fig:ECDF_GL_MT}
\end{figure}

From the aforementioned results, radar and positioning systems have the potential to outperform blind BM methods in each location within the coverage area. Moreover, they appear to be complementary technologies, suggesting that the integration (see Section~\ref{sec:MT}) of the two would provide the utmost BM performance. 
Figures~\ref{fig:training_time_MT} and~\ref{fig:ECDF_GL_MT} show $\overline{T}_{\mathrm{train}}$ and $\Delta \mathrm{G}$ of the optimal positioning system P1, the R2 radar configuration, and the fusion of these technologies (purple) according to~\eqref{eq:mt_variance}. 
P1 guarantees the minimum BM training time at the cost of an higher beamforming gain loss, mainly due to fast varying azimuth and elevation angles in time nearby the BS compared to the rate of acquisition of the positioning system ($R_\mathrm{pos}=10$ Hz). 
By contrast, radar-aided technique spends more time in beam training (more than twice as long as P1), mainly due to scatter point uncertainty, but maximizing the beamforming gain. 
As a matter of fact, combining these two approaches, leads to an improvement of the overall spectral efficiency. Indeed, the optimal P1 estimation of the angle, opportunely merged with the more fresh information of the radar, allows to minimize the gain losses at the cost of 0.2 ms more spent on $\Bar{T}_{\mathrm{train}}$. This latter aspect is manly due to the higher uncertainty of the extended target modelling.

\section{Concluding Remarks}\label{sect:conclusion}

Beam Management (BM) in high-frequency dynamic scenarios, such as V2X, presents several challenges that must be addressed to enable efficient and reliable communication systems. 
This paper offers a pragmatical comparison of different BM technologies in a realistic V2I urban scenario. Specifically, blind methods are compared with both position-assisted ones, leveraging positioning sensors at the VE side, e.g. GNSS, and radar-assisted ones, when the BS is equipped with a stand-alone radar. The comparison is quantified in terms of both beam training time (i.e., the delay introduced by the BM procedure), and beamforming gain loss (which depends on the accuracy of the beam selection). Interestingly, blind methods offer the best BM solution only when the VE continuously communicates with the BS. However, such a setup is not of practical interest in a vehicular network, as the BS shares its resources among multiple VEs yielding to discontinuous communications. Differently, there is not a winning technology for BM between positioning sensors at VE and radar at BS. Both provide complementary performance depending on the VE's location within the cell, suggesting the joint usage of both in a Multi-Technology BM approach. However, although radar-based BM does not require dedicated control signaling between BS and VE (as for position-based approaches), other factors limit the radar accuracy in localizing the VE's antenna. As demonstrated by the evidences of a dedicated experimental campaign, the back-scattering region from the vehicle is originated from a double bounce with the car body and the ground, biasing the estimation as the detected spot in the radar image is much larger than the resolution and generally not centered around the VE's antenna. This latter effect roughly doubles the required beam training time and leads to a 3 dB loss in beamforming gain in the considered settings, but the impact could be worse whenever the resolution of communication and/or sensing systems increase. Therefore, the accurate localization of the radar's antenna in 6G network sensing is still an open problem, whose solution would tip the scales toward the usage of radar-only BM.


\section*{Acknowledgment}

This research was carried out in the framework of the Huawei-Politecnico di Milano Joint Research Lab. The Authors want to acknowledge the Huawei Milan Research Centre.


\bibliographystyle{IEEEtran}
\bibliography{Bibliography}

\end{document}